\documentclass[11pt,a4paper]{elsarticle}
\usepackage[lining,scale=0.8]{FiraMono}

\usepackage[T1]{fontenc}

\usepackage{amsmath}
\usepackage{amssymb}
\usepackage{bm}
\usepackage{epsfig}
\usepackage{graphicx}
\usepackage[colorlinks=true]{hyperref}
\usepackage{listings}
\usepackage{ragged2e}
\usepackage{subcaption}
\usepackage{enumitem}

\input{all.tikzdefs}
\tikzstyle{blank}=[fill=white, shape=circle, draw=white, inner sep=0.8pt]
\tikzstyle{dot}=[fill=black, shape=circle, draw=black, inner sep=0.8pt]
\tikzstyle{fat}=[fill=white, shape=circle, draw={rgb,255: red,176; green,36; blue,39}, dashed, line width=1pt, inner sep=0.8pt]

\tikzstyle{edge}=[-, draw=EdgeColor, line width=1.2pt, line cap=rect, style=whitebg]
\tikzstyle{incoming edge}=[|-, style=edge, draw=PaleEdgeColor, style=arrowin]
\tikzstyle{outgoing edge}=[->, style=edge, draw=PaleEdgeColor, style=arrow]

\tikzstyle{massive edge}=[-, draw=MassiveEdgeColor, line width=2pt, style=whitebg, line cap=rect]
\tikzstyle{incoming massive edge}=[|-, style=massive edge, draw=PaleMassiveEdgeColor, style=arrowin]
\tikzstyle{outgoing massive edge}=[->, style=massive edge, draw=PaleMassiveEdgeColor, style=arrow]

\tikzstyle{fermion}=[-, draw=FermionColor, line width=1pt, style=whitebg, line cap=rect, postaction=decorate, decoration={markings,mark=at position .60 with {\arrow{stealth[round]}}}]
\tikzstyle{incoming fermion}=[-, style=fermion, draw=PaleEdgeColor]
\tikzstyle{outgoing fermion}=[-, style=fermion, draw=PaleEdgeColor]

\tikzstyle{massive fermion}=[-, draw=MassiveFermionColor, line width=1.9pt, line cap=rect, style=whitebg]
\tikzstyle{incoming massive fermion}=[-, style=massive fermion, draw=PaleEdgeColor]
\tikzstyle{outgoing massive fermion}=[-, style=massive fermion, draw=PaleEdgeColor]

\tikzstyle{gluon}=[-, draw=GluonColor, line width=1pt, preaction={{draw=white,line width=2pt}}, line cap=rect, decorate, decoration={coil,aspect=1.5,segment length=7pt}]
\tikzstyle{incoming gluon}=[-, style=gluon, draw=PaleEdgeColor]
\tikzstyle{outgoing gluon}=[-, style=gluon, draw=PaleEdgeColor]

\tikzstyle{ghost}=[-, style=fermion, line width=1pt, line cap=round, dash pattern={on 0pt off 3\pgflinewidth}]
\tikzstyle{incoming ghost}=[-, style=ghost, draw=PaleEdgeColor]
\tikzstyle{outgoing ghost}=[-, style=ghost, draw=PaleEdgeColor]

\tikzstyle{scalar}=[-, draw=ScalarColor, line width=1pt, densely dashed]
\tikzstyle{incoming scalar}=[-, style=scalar, draw=PaleScalarColor]
\tikzstyle{outgoing scalar}=[-, style=scalar, draw=PaleScalarColor]

\tikzstyle{massive scalar}=[-, draw=ScalarColor, line width=2pt, densely dashed]
\tikzstyle{incoming massive scalar}=[-, style=massive scalar, draw=PaleScalarColor]
\tikzstyle{outgoing massive scalar}=[->, style=massive scalar, draw=PaleScalarColor, style=arrow]

\tikzstyle{vector}=[-, draw=VectorColor, line width=1pt, preaction={{draw=white,line width=2pt}}, line cap=rect, decorate, decoration=snake]
\tikzstyle{incoming vector}=[-, style=vector, draw=PaleEdgeColor]
\tikzstyle{outgoing vector}=[-, style=vector, draw=PaleEdgeColor]

\tikzstyle{dot1}=[-, postaction=decorate, decoration={markings,mark=at position .50 with {\node[style=dot]{};}}]
\tikzstyle{dot2}=[-, postaction=decorate, decoration={markings,mark=between positions 0.33 and 0.67 step 0.33 with {\node[style=dot]{};}}]
\tikzstyle{dot3}=[-, postaction=decorate, decoration={markings,mark=between positions 0.25 and 0.76 step 0.25 with {\node[style=dot]{};}}]
\tikzstyle{dot4}=[-, postaction=decorate, decoration={markings,mark=between positions 0.20 and 0.81 step 0.20 with {\node[style=dot]{};}}]

\definecolor{grey}{RGB}{171,171,171}
\definecolor{EmeraldGreen}{HTML}{1ea78d}
\definecolor{EnglishRed}{HTML}{b02427}

\newcommand*{\sfig}[1]{{\scalebox{0.75}{{\input{#1}}}}}

\AtBeginDocument{\hypersetup{urlcolor=EmeraldGreen,citecolor=EmeraldGreen,linkcolor=EnglishRed}}

\setlist{parsep=\medskipamount}

\usepackage{prettyref}
\let\origref\ref
\newrefformat{chap}{\hyperref[#1]{Chapter~\origref*{#1}}}
\newrefformat{app}{\hyperref[#1]{Appendix~\origref*{#1}}}
\newrefformat{sec}{\hyperref[#1]{Section~\origref*{#1}}}
\newrefformat{subsec}{\hyperref[#1]{Section~\origref*{#1}}}
\newrefformat{tab}{\hyperref[#1]{Table~\origref*{#1}}}
\newrefformat{fig}{\hyperref[#1]{Figure~\origref*{#1}}}
\newrefformat{eq}{Eq.\,\hyperref[#1]{(\origref*{#1})}}
\AtBeginDocument{\renewcommand*{\ref}[1]{\prettyref{#1}}}

\usepackage{booktabs}
\usepackage{array}
\newcolumntype{M}[1]{>{\centering\arraybackslash}m{#1}}	% to center values in the table

\usepackage[super]{nth}

\usepackage[frozencache=true,cachedir=mintedcachedir,newfloat]{minted}

\definecolor{light-gray}{gray}{0.97}
\setminted{bgcolor=light-gray}
\usepackage[colorinlistoftodos, textsize=scriptsize]{todonotes}
\SetupFloatingEnvironment{listing}{name=Code Example}
\newcommand*\py{\lstinline[language=Python]}
\newcommand*\bash{\lstinline[language=Bash]}
\graphicspath{{plots/}}
\newcounter{bla}
\newenvironment{refnummer}{%
\list{[\arabic{bla}]}%
{\usecounter{bla}%
 \setlength{\itemindent}{0pt}%
 \setlength{\topsep}{0pt}%
 \setlength{\itemsep}{0pt}%
 \setlength{\labelsep}{2pt}%
 \setlength{\listparindent}{0pt}%
 \settowidth{\labelwidth}{[9]}%
 \setlength{\leftmargin}{\labelwidth}%
 \addtolength{\leftmargin}{\labelsep}%
 \setlength{\rightmargin}{0pt}}}
 {\endlist}
\def\be{\begin{equation}}
\def\ee{\end{equation}}
\def\bea{\begin{align}}
\def\eea{\end{align}}
\def\nn{\nonumber}

\newcommand{\pysecdec}{py{\textsc{SecDec}}}
\newcommand{\sympy}{{\textsc{sympy}}}

\newcommand{\eps}{\varepsilon}

\newcommand{\Disteval}{\textsc{Disteval}}

\usepackage[absolute]{textpos}
\AtBeginDocument{\begin{textblock*}{15cm}[1,0](16.8cm,3cm)\raggedleft\small\texttt{KA-TP-09-2023, P3H-23-035, IPPP/23/24}\end{textblock*}}

\begin{document}

\begin{frontmatter}

\title{Numerical Scattering Amplitudes with pySecDec}
% or: Speeding up amplitude evaluations in pySecDec

\author[a]{G.~Heinrich}
\author[b]{S.~P.~Jones}
\author[a]{M.~Kerner}
\author[a]{V.~Magerya}
\author[a]{A.~Olsson}
\author[c]{J.~Schlenk}

\address[a]{Institute for Theoretical Physics, Karlsruhe Institute of Technology (KIT),\\76128 Karlsruhe, Germany}
\address[b]{Institute for Particle Physics Phenomenology, Durham University,\\Durham DH1 3LE, UK}
\address[c]{ICS, University of Zurich, Winterthurerstrasse 190, \\8057 Zurich, Switzerland}
%\address[d]{Physik-Institut, Universit{\"a}t Z{\"u}rich, Winterthurerstrasse 190, 8057 Z{\"u}rich, Switzerland}

\begin{abstract}
We present a major update of the program \pysecdec{}, a toolbox for the evaluation of dimensionally regulated parameter integrals.
The new version enables the evaluation of multi-loop integrals as well as amplitudes in a highly distributed and flexible way, optionally on GPUs.
The program has been optimised and runs up to an order of magnitude faster than the previous release.
A new integration procedure that utilises construction-free median Quasi-Monte Carlo rules is implemented. 
The median lattice rules can outperform our previous component-by-component rules by a factor of 5 and remove the limitation on the maximum number of sampling points.
The expansion by regions procedures have been extended to support Feynman integrals with numerators, and functions for automatically determining when and how analytic regulators should be introduced are now available.
%For physics processes currently of interest, the new version is an of magnitude faster than the previous release.
The new features and performance are illustrated with several examples.
\end{abstract}

%\begin{flushleft}
%PACS: 12.38.Bx, %perturbative calculations
%02.60.Jh, 	%Numerical differentiation and integration
%02.70.Wz 	% Symbolic computation (computer algebra)
%\end{flushleft}


\end{frontmatter}

\newpage

{\bf PROGRAM SUMMARY}

\begin{small}
\noindent
{\em Manuscript Title: } Numerical Scattering Amplitudes with  py\textsc{SecDec}   \\
{\em Authors: } G.~Heinrich, S.~P.~Jones, M.~Kerner,
V.~Magerya, A.~Olsson, J.~Schlenk  \\
{\em Program Title: } py\textsc{SecDec}                                    \\
{\em Developer's repository:} \url{https://github.com/gudrunhe/secdec} \\
{\em Online documentation:} \url{https://secdec.readthedocs.io} \\
%{\em Journal Reference:}                           \\
%{\em Program Files doi:}                              \\
 {\em Licensing provisions: GNU Public License v3}                                   \\
  %enter "none" if CPC non-profit use license is sufficient.
{\em Programming language:} Python, \textsc{Form}, C++, \textsc{Cuda}     \\
{\em Computer:} from a single PC/Laptop to a cluster, depending on the
problem; if the optional GPU support is used, \textsc{Cuda} compatible hardware is required.\\
  %Computer(s) for which program has been designed.
{\em Operating system: } Unix, Linux                                      \\
  %Operating system(s) for which program has been designed.
{\em RAM:} hundreds of megabytes or more, depending on the complexity of the problem                                              \\
  %RAM in bytes required to execute program with typical data.
  %{\em Number of processors used:}                              \\
  %If more than one processor.
  %{\em Supplementary material:}                                 \\
  % Fill in if necessary, otherwise leave out.
 \\
%{\em PACS:}
%12.38.Bx,
%02.60.Jh, 	
%02.70.Wz \\
{\em Classification:}
  4.4 Feynman diagrams,
  5 Computer Algebra,
  11.1 General, High Energy Physics and Computing.\\
{\em External routines/libraries:}
\textsc{GSL}~[1],
\textsc{NumPy}~[2],
\textsc{SymPy}~[3],
\textsc{Nauty}~[4],
\textsc{Cuba}~[5],
\textsc{Form}~[6],
\textsc{GiNaC} and \textsc{CLN}~[7],
\textsc{Normaliz}~[8],
\textsc{GMP}~[9]. \\
%{\em Subprograms used:}                                       \\
%Fill in if necessary, otherwise leave out.
%{\em Catalogue identifier of previous version:}*              \\
%Only required for a New Version summary, otherwise leave out.
{\em Journal reference of previous version:} \href{https://doi.org/10.1016/j.cpc.2021.108267}{Comput. Phys. Commun. 273 (2022) 108267}~\cite{Heinrich:2021dbf}.\\
{\em Does the new version supersede the previous version?:} yes  \\
{\em Nature of the problem:}\\
 Scattering amplitudes at higher orders in perturbation theory are
 typically represented as a linear combination of coefficients ---
 containing the kinematic invariants and the space-time dimension ---
 multiplied with loop integrals which contain singularities and whose analytic representation might be
 unknown. \\
  {\em Solution method:}\\
  Extraction of singularities in the dimensional regularization
 parameter as well as in analytic regulators for potential spurious
 singularities is done using sector decomposition.
 The combined evaluation of the integrals with their coefficients is
 performed in an efficient way.
    \\
%{\em Reasons for the new version:}*\\
%{\em Summary of revisions:}*\\
{\em Restrictions:} Depending on the complexity of the problem, limited by
memory and CPU/GPU time.\\
%   \\
{\em Running time:}
Between a few seconds and several days, depending on the comp\-lexity of the problem.\\
  %Give an indication of the typical running time here.
{\em References:}
\begin{refnummer}
\item M.~Galassi et al,
    GNU Scientific Library Reference Manual.
    \texttt{ISBN:0954612078},
    \url{http://www.gnu.org/software/gsl/}.
\item C.~R.~Harris, K.~J.~Millman, S.~J.~van~der~Walt, et al,
    Array programming with NumPy,
    Nature \textbf{585} (2020) 357--362.
    \texttt{\href{https://doi.org/10.1038/s41586-020-2649-2}{doi:10.1038/s41586-020-2649-2}},
    \url{http://www.numpy.org/}.
\item A.~Meurer, et al.,
    SymPy: symbolic computing in Python,
    PeerJ\ Comp.\ Sci. \textbf{3} (2017) e103.
    \texttt{\href{https://doi.org/10.7717/peerj-cs.103}{doi:10.7717/peerj-cs.103}},
    \url{http://www.sympy.org/}.
\item B.~D.~McKay and A.~Piperno,
    Practical graph isomorphism, II,
    J.\ Symb.\ Comput.\ \textbf{60} (2014) 94--112.
    \texttt{\href{https://doi.org/10.1016/j.jsc.2013.09.003}{doi:10.1016/j.jsc.2013.09.003}},
    \url{http://pallini.di.uniroma1.it}.
\item T.~Hahn,
    CUBA: A Library for multidimensional numerical integration,
    Comput.\ Phys.\ Commun.\ \textbf{168} (2005) 78.
    \texttt{\href{https://arxiv.org/abs/hep-ph/0404043}{arXiv:hep-ph/0404043}},
    \url{http://www.feynarts.de/cuba/}.
\item J.~Kuipers, T.~Ueda and J.~A.~M.~Vermaseren,
    Code Optimization in FORM,
    Comput.\ Phys.\ Commun.\ \textbf{189} (2015) 1.
    \texttt{\href{https://arxiv.org/abs/1310.7007}{arXiv:1310.7007}},
    \url{http://www.nikhef.nl/~form/}.
\item C.~W.~Bauer, A.~Frink, and R.~B.~Kreckel,
    Introduction to the GiNaC framework for symbolic computation within the C++ programming language,
    J.\ Symb.\ Comput.\ \textbf{33} (2002) 1--12.
    \texttt{\href{https://arxiv.org/abs/cs/0004015}{arXiv:cs/0004015}},
    \url{https://www.ginac.de/}.
\item W.~Bruns, B.~Ichim, B. and T.~R{\"o}mer, C.~S{\"o}ger,
    Normaliz. Algorithms for rational cones and affine monoids.
    \url{http://www.math.uos.de/normaliz/}.
\item T.~Granlund et al,
    GMP: The GNU Multiple Precision Arithmetic Library.
    \url{https://gmplib.org/}.
\end{refnummer}
\end{small}

%\hspace{1pc}
%{\bf LONG WRITE-UP}

\section{Introduction}
\label{sec:intro}

The calculation of scattering amplitudes beyond one loop is required in order to provide predictions for the increasingly precise measurements at the LHC, at B-factories and at other colliders.
Furthermore, future lepton colliders require substantial progress in the calculation of higher order electroweak corrections, which usually involve several mass scales. The latter pose challenges for the evaluation of the corresponding integrals, in particular for analytic approaches.
The program (py)\textsc{SecDec}~\cite{Borowka:2012yc,Borowka:2015mxa,Borowka:2017idc,Borowka:2018goh} offers the possibility to calculate multi-scale integrals beyond one loop numerically.
Other public programs for the numerical evaluation of multi-loop integrals based on sector decomposition within dimensional regularisation~\cite{Binoth:2000ps,Heinrich:2008si} are \texttt{sector\_decomposition}~\cite{Bogner:2007cr} and \textsc{Fiesta}~\cite{Smirnov:2008py,Smirnov:2009pb,Smirnov:2013eza,Smirnov:2015mct,Smirnov:2021rhf}. The program \textsc{Feyntrop}~\cite{Borinsky:2023jdv} provides a numerical approach for evaluating quasi-finite Feynman integrals using tropical sampling~\cite{Borinsky:2020rqs}.
Other analytic/semi-analytic approaches include \textsc{DiffExp}~\cite{Hidding:2020ytt,Moriello:2019yhu},  \textsc{AMFlow}~\cite{Liu:2022chg} and \textsc{SeaSyde}~\cite{Armadillo:2022ugh} which calculate Feynman integrals by solving differential equations using series expansions.

%%%%%%%%%

The program \pysecdec{} has been upgraded recently with the ability to perform expansions by regions~\cite{Heinrich:2021dbf}, a method pioneered in Refs.~\cite{Smirnov:1991jn,Beneke:1997zp,Pak:2010pt,Jantzen:2011nz}.
Ref.~\cite{Heinrich:2021dbf} also describes an early implementation of an algorithm for efficiently calculating the weighted sum of integrals.

In this paper, we present \pysecdec{} version 1.6, which is a major upgrade in several respects.
One of the main changes is the fact that  much more general coefficients of integrals than previously  allowed
are now supported. This feature is important for the calculation of amplitudes in a form resulting from IBP reduction, where the coefficients of the master integrals are usually sums of large rational polynomials containing kinematic invariants and the space-time dimension $D$.
Furthermore, various changes in the code structure and numerical evaluation lead to a significant speed-up of the numerical evaluation. We present a new Quasi-Monte-Carlo (QMC) evaluator, called \textsc{Disteval}, which is optimised for a highly distributed evaluation. Another major improvement is achieved by the use of median generating vectors for the rank-1 lattice rules the QMC integration is based on.
In addition, the feature of expansion by regions has been upgraded. For example, the program can automatically detect whether a regulator in addition to the dimensional regulator is needed in certain regions. In addition, the algebraic expressions multiplying each order of the expansion in a small parameter are provided to the user.

This article is structured as follows. In \ref{sec:newfeatures} the new features of version 1.6 are described. In \ref{sec:examples} we present examples which demonstrate the usage of the program and the new features, as well as timings comparing previous \pysecdec{} versions to the current version. Conclusions are presented in \ref{sec:conclusion}.

The release version of the code is available at \url{https://pypi.org/project/pySecDec/} and can be obtained via \textsc{pip}.
The development version lives at \url{https://github.com/gudrunhe/secdec}.
Online documentation can be found at \url{https://secdec.readthedocs.io/}.

\section{New features of \pysecdec}
\label{sec:newfeatures}

The main new features of \pysecdec{} version 1.6 are a new integrator/importance sampling procedure (\Disteval{}), support for construction-free median Quasi-Monte Carlo rules and improved support for expansion by regions.

The \Disteval{} integrator is presented in~\ref{sec:disteval}, it implements a newly constructed Quasi-Monte-Carlo (QMC) integrator and is significantly faster and more configurable than our previous integrators. 
The \Disteval{} integrator also comes with much better support for inputting complicated coefficients of the master integrals, including sums of rational functions resulting from the IBP reduction of amplitudes.

In~\ref{sec:medianlattice}, we describe and provide benchmarks of our implementation of \textit{median Quasi-Monte Carlo rules}, a new QMC lattice construction based on Ref.~\cite{medianQMC}.
The median QMC rules are made available in the \textsc{Qmc} and the \Disteval{} integrators.

Improvements to the expansion by regions routines are described in~\ref{sec:extraregulators}. 
The new version of \pysecdec{} supports Feynman integrals with numerators and provides functions for determining where an additional extra regulator, in addition to dimensional regularisation, is needed.

%Several examples demonstrating the performance of this release and the new features are given in~\ref{sec:examples}.

\subsection{The new Quasi-Monte-Carlo evaluator \textsc{Disteval}}
\label{sec:disteval}

\pysecdec{} traditionally comes with support for multiple
integrators: \texttt{Qmc} based on the \textsc{Qmc}
library~\cite{Borowka:2018goh}; \texttt{Vegas}, \texttt{Suave},
\texttt{Divonne}, and \texttt{Cuhre} based on the \textsc{Cuba}
library~\cite{Hahn:2016ktb}; \texttt{CQuad} based on the
\textsc{GSL} library~\cite{GMP}.
Out of these we have recommended the usage of the \texttt{Qmc}
integrator as the only one that achieves super-linear scaling
of the integration precision with integration time for practical
multidimensional integrals.
All of these six integrators are available through a unified
integration interface we shall call ``\textsc{IntLib}'' (for
lack of a better name).

With the new version of \pysecdec{} we introduce a new integration
interface and an integrator ``\textsc{Disteval}''.
\textsc{Disteval} implements a Randomized Quasi-Monte-Carlo (RQMC)
integration method based on rank-1 shifted lattice rules~\cite{QMCActaNumerica,Li:2015foa}.
It is directly analogous to the \textsc{IntLib} \texttt{Qmc}
integrator, but with significantly higher performance, and the
possibility of evaluation distributed across several computers.
As with \texttt{Qmc}, \textsc{Disteval} supports both CPUs and
GPUs, with the latter ones being preferred due to their speed.

In \ref{sec:examples} we provide a series of benchmarks demonstrating
the speedup \textsc{Disteval} provides over \texttt{Qmc}
(usually between 3x and 10x)
across a variety of integrals, on both CPUs and GPUs.
%, all while having the same convergence properties.

There are multiple sources of this speedup:

\begin{itemize}
\item
    While \textsc{IntLib} integrands are compiled separately
    from the integration algorithms and are called indirectly by
    the integrators, \textsc{Disteval} integrands fully include
    the integration loop.
    This enables the hoisting of the common code from the integration
    loop, the fusion of the lattice point generation and the
    integrand evaluation, and multiple micro-optimizations by
    the compiler.
    This however comes at the expense of flexibility in choosing
    integrators.
\item
    The code for GPU integrands and CPU integrands are generated
    separately, allowing for separate optimization to be applied
    for each.
\item
    On the GPU side \textsc{Disteval} uses the highly optimized
    NVidia CUB library\footnote{\url{https://github.com/NVIDIA/cub}}
    to sum up the samples on the GPU (instead of performing the
    sum on the CPU), minimizing the data transfer between CPU
    and GPU.
\item
    Modern CPUs are capable of executing multiple independent
    instructions in parallel.
    For example, an AMD~\textsc{Epyc}~7F32 processor contains four
    floating-point execution units: two capable of performing one
    256-bit Fused Multiply-Add (FMA) operation per cycle each,
    and two capable of one 256-bit addition operation per cycle
    each, for the total of 16 double-precision (i.e. 64-bit)
    operations per cycle.
    Saturating these executing units with work is essential in
    achieving optimal performance, and the best way to do that
    is to structure the code to operate on multiple values at
    the same time, packing 64-bit double-precision values into
    256-bit arrays and utilizing SIMD\footnote{``Single Instruction
    Multiple Data.''} instructions that operate on the whole
    array at once.

    The integrand kernels \pysecdec{} generates for \textsc{Disteval}
    do exactly this: each mathematical operation is coded to
    work on 4 double-precision values simultaneously, and if the
    compiler is allowed to emit 256-bit SIMD instructions (i.e.
    via the AVX2 and FMA instructions sets on x86 processors),
    each such operation becomes a single instruction.

    Note that while all modern x86 processors support AVX2 and FMA,
    some older ones do not, and because of this \textsc{Disteval}
    does not require their support.
    It is up to the user to check if all their target machines
    have this support,\footnote{This can be done by checking
    the presence of \texttt{avx2} and \texttt{fma} flags in
    \texttt{/proc/cpuinfo}.} and if so, to allow the compiler to
    use these instruction sets by e.g. setting \texttt{CXXFLAGS}
    to \texttt{-mavx2~-mfma} during compilation.\footnote{See e.g.
    \url{https://gcc.gnu.org/onlinedocs/gcc/x86-Options.html}
    for a description of machine-specific options of GCC.}
    This is highly recommended.

    Users that plan to perform integration on a single machine are
    advised to set \texttt{CXXFLAGS} to \texttt{-march=native},
    so that the compiler would be allowed to auto-detect the
    capabilities of the processor it is running on, and use all
    the available instruction sets.
\item
    Multiple smaller micro-optimizations on the CPU and the GPU
    sides to reduce the overhead for smaller integrands, and to
    speed up larger ones.
\end{itemize}

\subsubsection{Using \textsc{Disteval}}

Usage-wise, \textsc{Disteval} diverges from \textsc{IntLib},
during compilation and integration, but is similar enough that
porting integration scripts should be easy.

As an example, let us consider a massless one-loop box.
To generate the integration library for both integration interfaces,
one can use the following Python script:

\begin{lstlisting}[language=python]
import pySecDec as psd
if __name__ == "__main__":
    li = psd.LoopIntegralFromPropagators(
        loop_momenta=["l"],
        external_momenta=["p1","p2","p3"],
        propagators=["l**2","(l-p1)**2","(l-p1-p2)**2","(l-p1-p2-p3)**2"],
        replacement_rules=[
            ("p1*p1","0"),   ("p2*p2","0"),   ("p3*p3","0"),
            ("p1*p2","s/2"), ("p2*p3","t/2"), ("p1*p3","-s/2-t/2")])
    psd.loop_package(
        name="box1L",
        loop_integral=li,
        real_parameters=["s","t"],
        requested_orders=[0])
\end{lstlisting}

Then, to compile the \textsc{IntLib} library one can invoke
\bash{make} from the command shell:
\begin{lstlisting}[language=bash]
make -C box1L -j4
\end{lstlisting}

Similarly, to compile the \textsc{Disteval} library one can
use:\footnote{As noted earlier, adding \bash{CXXFLAGS="-mavx2
-mfma"} to this \bash{make} call is recommended.}
\begin{lstlisting}[language=bash]
make -C box1L -j4 disteval
\end{lstlisting}

The resulting library will be fully contained in the
\texttt{box1L/disteval/} directory, meaning that the directory
can be freely moved to a different location.
The file \texttt{box1L/disteval/box1L.json} will contain the
full description of the requested integral, and will work as the
entry point to the library.

If one wants to use the resulting library on a GPU with ``compute
capability''~8.0, one should add \bash{SECDEC_WITH_CUDA_FLAGS="-arch=sm_80"}
to the arguments of the \bash{make} call.\footnote{The list
of NVidia ``Compute Capability'' codes for different GPUs is
available at \url{https://developer.nvidia.com/cuda-gpus}.}
For \textsc{IntLib} this will build a library that can only be
used on the GPU; for \textsc{Disteval} the resulting library
will be able to work with and without a GPU.

To integrate using \textsc{IntLib} one can use the Python
interface:
\begin{lstlisting}[language=python]
from pySecDec.integral_interface import IntegralLibrary
lib = IntegralLibrary("box1L/box1L_pylink.so")
lib.use_Qmc()
_, _, result = lib(real_parameters=[4.0, -0.75], epsrel=1e-3, epsabs=1e-8)
print(result)
\end{lstlisting}

Similarly, to integrate using \textsc{Disteval} one can use the
Python interface:
\begin{lstlisting}[language=python]
from pySecDec.integral_interface import DistevalLibrary
lib = DistevalLibrary("box1L/disteval/box1L.json")
result = lib(parameters={"s": 4.0, "t": -0.75}, epsrel=1e-3, epsabs=1e-8)
print(result)
\end{lstlisting}

Alternatively, one can also use the new command-line interface:
\begin{lstlisting}[language=bash]
python3 -m pySecDec.disteval box1L/disteval/box1L.json \
    s=4 t=-0.75 --epsrel=1e-3 --epsabs=1e-8
\end{lstlisting}

\subsubsection{Distributed evaluation}

The integrand evaluation under \textsc{Disteval} is performed
by worker processes, while the main process is responsible for
distributing work among the workers and processing the results.
Communication between the main and the worker processes is done
via bidirectional bytestreams (i.e. pipes), using a custom
\texttt{json}-based protocol, which means that the workers do not need
to be located on the same machine as the main process.

By default, the Python interface of \textsc{Disteval} will launch
one worker process per locally available GPU, or one per locally
available CPU.
Each CPU worker is launched with the command

\begin{lstlisting}[language=bash]
python3 -m pySecDecContrib pysecdec_cpuworker
\end{lstlisting}

and each GPU worker is launched with the command

\begin{lstlisting}[language=bash]
python3 -m pySecDecContrib pysecdec_cudaworker -d <i>
\end{lstlisting}

where \bash{<i>} is the (zero-based) index of the GPU this worker
should use.

The default worker selection however can be overridden through the
\py{workers} argument of \py{DistevalLibrary} to allow execution
on different machines.
For example, suppose that the integration is to be spread across
two machines: \texttt{gpu1} with a single GPU, and \texttt{gpu2}
with two GPUs; if both machines are reachable via \texttt{ssh}, then
one could setup the integration library as follows:
\begin{lstlisting}[language=python]
lib = DistevalLibrary(
    "box1L/disteval/box1L.json",
    workers=[
        "ssh gpu1 python3 -m pySecDecContrib pysecdec_cudaworker -d 0",
        "ssh gpu2 python3 -m pySecDecContrib pysecdec_cudaworker -d 0",
        "ssh gpu2 python3 -m pySecDecContrib pysecdec_cudaworker -d 1"
    ])
\end{lstlisting}

\subsubsection{Adaptive weighted sum evaluation}

Since \pysecdec{} version~1.5 \textsc{IntLib} supports adaptive
integration of weighted sums of integrals (e.g. amplitudes) via
the \py{sum_package()} function.
Versions of \py{loop_package()} and \py{make_package()}
implemented in terms of \py{sum_package()} also have been added.
\textsc{Disteval} implements a very similar adaptive sampling
algorithm.

Suppose we have a set of integrals $I_i$, and we want to calculate
a set of their weighted sums $A_k\equiv\sum_i C_{ki}I_i$.
When evaluated under RQMC, each $I_i$ can be thought of as a
normally distributed random variable,
\begin{equation}
    I_i\sim\mathcal{N}(\mathrm{mean}(I_i),\mathrm{var}(I_i)).
\end{equation}

Let us assume that it takes $\tau_i$ of time to evaluate the
integrand of $I_i$ once, and that $\mathrm{var}(I_i)$ scales
with the number of integrand evaluations $n_i$ (a.k.a. the size
of the lattice on which the integrand is evaluated) as
\begin{equation}
    \mathrm{var}(I_i)=\frac{w_i}{n_i^{\alpha}}.
    \label{eq:varscaling}
\end{equation}
Our objective then is to choose $n_i$ as functions of $C_{ki}$,
$w_i$, $\tau_i$, and $\alpha$, to minimize the total integration
time
\begin{equation}
    T \equiv \sum_i \tau_i n_i,
\end{equation}
while achieving the total variance $V_k$ requested by the user:
\begin{equation}
    \mathrm{var}(A_k)=\sum_i \left| C_{ji} \right|^2 \frac{w_i}{n_i^{\alpha}} = V_k \; \left( \forall k \right).
\end{equation}

We solve this optimization problem via the Lagrange multiplier
method:
\begin{equation}
    L \equiv T+\sum_k \lambda_k \left( \mathrm{var}(A_k) - V_k \right),\qquad\text{and}\qquad\frac{\partial L}{\partial \left\{ n_i,\lambda_k\right\} }=0.
\end{equation}

If only one sum $A_k$ needs to be evaluated, then these equations have a closed-form solution:
\begin{equation}
    \begin{split}
        \lambda_k &= \frac{1}{\alpha}\left(\frac{1}{V_k}\sum_{k}\left(\left|C_{jk}\right|^{2}w_{k}\tau_{k}^{\alpha}\right)^{\frac{1}{\alpha+1}}\right)^{\frac{\alpha+1}{\alpha}}\!, \\
        n_i &= \left(\frac{\alpha w_i}{\tau_i}\lambda_k\left|C_{ji}\right|^{2}\right)^{\frac{1}{\alpha+1}}\!.
    \end{split}
    \label{eq:adaptive-solution}
\end{equation}

If multiple sums are requested, \textsc{Disteval} uses this
formula first for the first sum, then updates $n_i$ and applies
it to the next sum, and so on.
%This procedure is not guaranteed to produce the optimal $n_i$
%assignments, but it does work in practice.

To make this work in practice, \textsc{Disteval} needs to estimate
the integral evaluation speed $\tau_i$, convergence constants
$w_i$, and the power $\alpha$.
The evaluation speed $\tau_i$ is estimated on-line, by first
benchmarking the relative performance of each worker, and then
by tracking how fast a given integral is being evaluated on a
given worker.
The convergence constants $w_i$ are first estimated by evaluating
all integrals with some preset minimum lattice size ($10^4$ by
default), and then updated each time an integration result is
obtained.
The parameter $\alpha$ is chosen conservatively to be~2, which
is the minimum asymptotic scaling guaranteed by the use of QMC
methods (for some examples see \ref{fig:hexatriangle-timing-plot}
where $\alpha\approx3$, and \ref{fig:pentabox-offshell-timing-plot}
where $\alpha\approx2$).

Here it is important to note that the scaling law of \ref{eq:varscaling}
is only asymptotic.
In practice the usage of rank-1 lattice rules means that for each
lattice size~$n_i$ we must construct a completely new lattice,
and often larger~$n_i$ results in a larger error, instead of a
smaller one --- a phenomenon which we call \textit{unlucky lattices}.

\begin{figure}
  \centering
  \includegraphics[width=\textwidth]{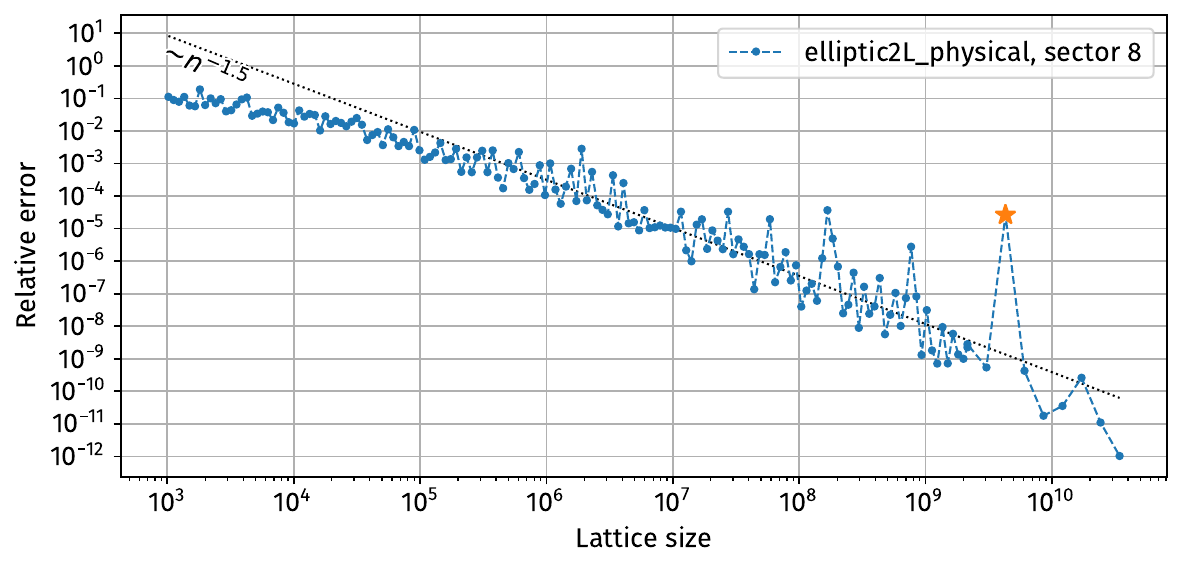}
  \includegraphics[width=\textwidth]{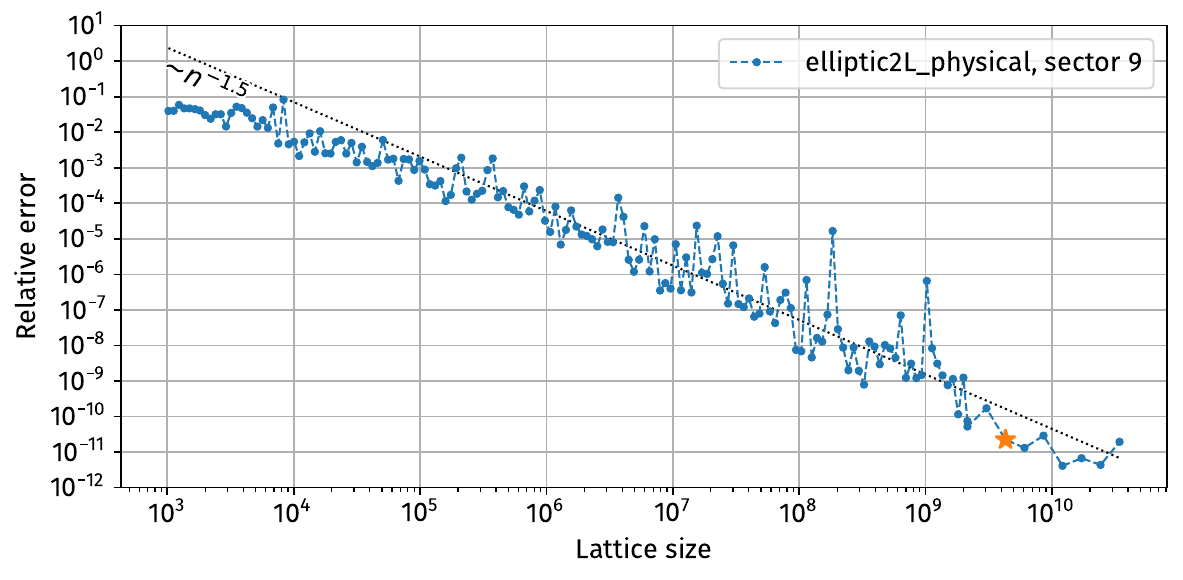}
  \caption{
      The RQMC integration error (i.e. $\sqrt{\textrm{var}(I_i)/m}$)
      after $m=32$ repetitions for lattices of different sizes.
      The integrals are sectors of the \texttt{elliptic2L\_physical}
      example from \ref{sec:timings}.
      The lattices are taken from the \textsc{Qmc} library, and
      are the same for both integrals.
      The result of one particularly unlucky lattice is marked
      with a star; note that this lattice is only unlucky for one
      of the sectors and performs normally for the others.
  }
  \label{fig:precision-by-lattice}
\end{figure}

As an illustration, consider \ref{fig:precision-by-lattice}:
although the variance overall scales as $1/n^3$ (and thus the
error as $1/n^{1.5}$), the progression is not monotonic, and
one particularly unlucky lattice results in an integration
error more than four orders of magnitude worse than lattices of
similar size around it --- but only for one of the integrals, for
the other the same lattice gives a perfectly good result.

This scaling structure makes the integration times inherently
unpredictable: if during the integration an integral is evaluated
on an unlucky lattice, then \textsc{Disteval} will overestimate
the integral's $w_i$ parameter, and will assume that many more
samples of this integral are needed to achieve the requested
precision, wasting integration time.
The practical impact of this is usually low to moderate, unless
one encounters a very unlucky lattice such as the one marked
with a star in \ref{fig:precision-by-lattice}.
To some extent, this effect can be tamed by the \textit{median QMC rules}, introduced in the following section.

\subsection{Median Quasi-Monte Carlo rules}
\label{sec:medianlattice}
The Quasi-Monte Carlo integration in previous versions of \pysecdec{} was based on pre-computed generating vectors, provided with the \textsc{Qmc} library~\cite{Borowka:2018goh}. These generating vectors were constructed using the component-by-compo\-nent~(CBC) method~\cite{CBC}, minimizing the worst-case error of the QMC integration, assuming arbitrary integrands belong to a Korobov space with smoothness $\alpha=2$ and using product weights.

However, for a given integrand, a lattice of size $n$ based on
the above CBC~construction might not be the optimal choice,
resulting in the unlucky lattices illustrated in the previous
section.
Furthermore, constructing lattices via the CBC method is
computationally expensive, and the largest such lattice currently
provided by the \textsc{Qmc} library has $\sim 7\cdot 10^{10}$
sampling points.
If the requested precision of the integral can not be achieved with
the largest available lattice, the error can only be improved by
repeated sampling of this lattice with random shifts, resulting
in a $n^{-1/2}$ scaling of the integration error, negating the
benefits of QMC integration.

\begin{table}
    \centering
    \scalebox{0.85}{%
        \begin{tabular}{lrrrrrrrrr}
        \toprule
        \textsubscript{Lattices}\textbackslash\textsuperscript{Accuracy}
            & $10^{-2}$ & $10^{-3}$ & $10^{-4}$ & $10^{-5}$ & $10^{-6}$ & $10^{-7}$ & $10^{-8}$ & $10^{-9}$ & $10^{-10}$ \\
        \midrule
        CBC            & 1.7\,s & 1.8\,s & 1.8\,s & 2.3\,s & 3.9\,s & 18\,s & 452\,s & 51.6\,m & 98.9\,m \\
        Median, $r=3$  & 1.7\,s & 1.7\,s & 1.8\,s & 2.2\,s & 3.7\,s & 12\,s & 44\,s & 3.3\,m & 13.8\,m \\
        Median, $r=5$  & 1.7\,s & 1.7\,s & 1.8\,s & 2.2\,s & 3.7\,s & 12\,s & 44\,s & 2.8\,m & 8.0\,m \\
        Median, $r=7$  & 1.7\,s & 1.8\,s & 1.7\,s & 2.1\,s & 4.2\,s & 12\,s & 39\,s & 2.8\,m & 9.4\,m \\
        Median, $r=11$ & 1.7\,s & 1.7\,s & 1.8\,s & 2.2\,s & 3.7\,s & 12\,s & 37\,s & 2.6\,m & 7.5\,m \\
        Median, $r=15$ & 1.7\,s & 1.8\,s & 1.8\,s & 2.2\,s & 3.5\,s & 10\,s & 38\,s & 2.8\,m & 8.2\,m \\
        Median, $r=23$ & 1.7\,s & 1.8\,s & 1.9\,s & 2.3\,s & 3.9\,s & 12\,s & 39\,s & 2.7\,m & 14.8\,m \\
        Median, $r=31$ & 1.7\,s & 1.9\,s & 2.0\,s & 2.4\,s & 4.3\,s & 14\,s & 46\,s & 3.5\,m & 11.1\,m \\
        Median, $r=63$ & 1.8\,s & 2.0\,s & 2.2\,s & 2.9\,s & 5.8\,s & 21\,s & 66\,s & 4.6\,m & 16.7\,m \\
        \bottomrule
        \end{tabular}%
    }
    \caption{
        Average integration times for the \texttt{elliptic2L\_physical}
        example using the \textsc{Disteval} integrator depending
        on the requested accuracy and the lattice construction
        method, comparing lattices derived via CBC and median QMC rules. The timings were taken using an NVidia~A100~80G GPU,
        with the integrands compiled using \textsc{Cuda}~11.8.89.
        \label{tab:elliptic2L_physical-median-timings}
    }
\end{table}

An alternative to the CBC construction called \textit{median QMC rules} has been proposed in~\cite{medianQMC}.
This construction is based on the observation that most generating vectors are good choices, provided the components are chosen from the set
\begin{align}
  \mathbb U_n \in \{1 \le z \le n-1 \,|\,\mathrm{gcd}(z,n)=1\}.
\end{align}
For $r$ randomly selected generating vectors $\mathbf z_1,\ldots,\mathbf z_r$ satisfying this condition, it has been shown that using the median
\begin{align}
  M_{n,r}(f) = \mathrm{median}(Q_{n,\mathbf z_1}(f),\ldots,Q_{n,\mathbf z_r}(f))
  \label{eq:median}
\end{align}
as an integral estimate results in the same convergence rate as the CBC construction with high probability (the larger~$r$ is chosen, the higher the probability).
Here, $Q_{n,\mathbf z}(f)$ is the estimate for the integral of $f$, obtained using the rank-1 lattice rule with generating vector~$\mathbf z$.

\begin{figure}
  \centering
  \includegraphics[width=\textwidth]{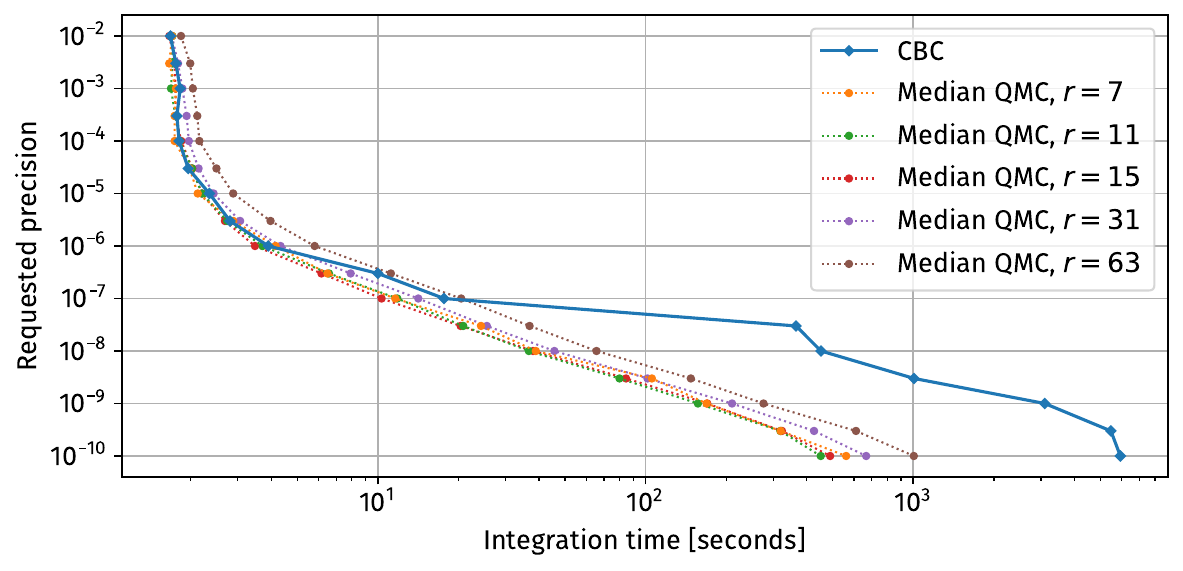}
  \caption{
      Integration time of the \texttt{elliptic2L\_physical}
      example from \ref{sec:timings} using the median QMC rules compared to the integration using CBC construction of the generating vectors.
      This plot uses the same benchmarking setup as \ref{tab:elliptic2L_physical-median-timings}.
  }
  \label{fig:medianQmc}
\end{figure}

In \pysecdec{} we now provide the possibility for an automated construction of generating vectors following this method. It can be enabled with the option \texttt{lattice\_candidates=r}, which specifies the number $r$ of randomly chosen generating vectors. 
After selecting the generating vector according to the median QMC rules, the uncertainty of the integration is then obtained by sampling the integrand on $m$ different random shifts of this lattice, as in previous versions of \pysecdec{}. Using this method, the construction of lattices of arbitrary size $n$ is possible, and since the generating vectors are chosen individually for each integrand, the problems due to unlucky lattices becomes less pronounced. With the default setting \texttt{lattice\_candidates=0}, only the pre-computed generating vectors based on CBC construction are used. 

As an illustration, consider the integration time of the \texttt{elliptic2L\_physical} example presented in \ref{tab:elliptic2L_physical-median-timings} and \ref{fig:medianQmc}.
With CBC lattices, two of the most complicated sectors of the example share a particularly unlucky lattice at $n = 4.3\cdot\!10^9$, as depicted in \ref{fig:precision-by-lattice}.
Starting with the requested precision of around $3\cdot\!10^{-8}$ the evaluator consistently hits these lattices, and the integration time goes up significantly.
On the other hand, the integration time with median QMC rules increases smoothly across the whole range.

A disadvantage of the median QMC rules is that, compared to using pregenerated lattices, an extra $r$ samples of the integral are required in addition to the $m$ samples used to estimate the integral uncertainty.
In practice we typically find that despite this overhead, the integration time using the median QMC rules is either comparable~to or improves upon that using lattices pregenerated via the CBC construction.
This of course depends on the number of lattice candidates, $r$: with small $r$ unlucky lattices are more likely, while large $r$ means more overhead.
For the \texttt{elliptic2L\_physical} example, this behavior can be seen in \ref{tab:elliptic2L_physical-median-timings}: $r=11$ and $r=15$ seem to perform the best overall, while lower and higher $r$ result in more integration time on average.

We have tested applying the generating vectors obtained for one particular integral using the median QMC rules to different integrals, thus lowering the overhead by avoiding the construction of new generating vectors for each integrand. However, we find that this typically leads to longer integration times, as a median lattice selected for a given integrand often does not turn out to be a high-quality choice for other integrands.

\subsection{Extra regulators for Expansion by Regions}
\label{sec:extraregulators}

When expansion by regions is applied to a well-defined dimensionally regulated integral, new spurious singularities may be introduced which are not regulated by the original regulator.
It is possible to detect geometrically which integrals will become ill-defined after expansion~\cite{Heinrich:2021dbf}.

One way to handle the new singularities is to generalise the definition of the integral by adding new analytic regulators, $\delta_1, \ldots, \delta_N$.
Commonly, this is done for Feynman integrals by altering the power of Feynman propagators according to $(\nu_1 \rightarrow \nu_1 + \delta_1, \ldots, \nu_N \rightarrow \nu_N + \delta_N)$, or, in the Feynman parametrisation, by multiplying the integrand by $x_1^{\delta_1} \cdots x_N^{\delta_N}$, where $x_i$ are Feynman parameters.
Introducing $N$ independent new regulators can dramatically increase the complexity of the problem and is often unnecessary.
Using the algorithms described in Ref.~\cite{Heinrich:2021dbf}, several new routines for detecting and handling spurious divergences have been added to \pysecdec{}, focusing on Feynman (loop) integrals.

The \texttt{loop\_regions} function now accepts the argument  \texttt{extra\_regulator\_name}.
If a string or symbol is passed to this argument, \pysecdec{} automatically determines if an extra regulator is required and, if so, introduces a single new regulator.
The integrand is multiplied by  $\mathbf{x}^{\bm{\nu}_\delta \delta}$ where $\delta$ is the extra regulator and $\bm{\nu}_\delta$ is a vector of integers automatically chosen such that the integral becomes well-defined.
Alternatively, the user may pass a specific $\bm{\nu}_\delta$ as a list of integers or \sympy{} rationals via the argument \texttt{extra\_regulator\_exponent}.

The function \texttt{suggested\_extra\_regulator\_exponent}, which the user can call independently of \texttt{loop\_regions}, automatically determines a vector of integers $\bm{\nu}_\delta$ sufficient to make a loop integral well-defined.
Given a \texttt{loop\_integral} object and the parameter in which it should be expanded, \texttt{smallness\_parameter}, the function returns $\bm{\nu}_\delta$.
There is considerable freedom in choosing the entries of $\bm{\nu}_\delta$.
The only important property is that its entries must obey a set of inequalities which ensure it is not tangent to any of the hyperplanes spanned by the set of new (internal) facets, introduced by the expansion, which lead to spurious singularities.
The \texttt{suggested\_extra\_regulator\_exponent} function returns only one choice for $\bm{\nu}_\delta$, it obeys the additional constraint $\sum_i \bm{\nu}_{\delta,i} = 0$, which ensures that the new regulator does not appear in the power of the $\mathcal{U}$ or $\mathcal{F}$ polynomials.

The function \texttt{extra\_regulator\_constraints} provides the list of constraints which must be obeyed by the entries of $\bm{\nu}_\delta$ for it to regulate the new singularities.
The user may call this function independently, for example, if they wish to impose additional constraints on the analytic regulators or if they want to understand the regions giving rise to spurious singularities and how they cancel.
The function returns a dictionary of regions and constraints that must be obeyed in order to obtain regulated integrals, along with a complete list of all constraints (the \texttt{all} entry).
Each set of constraints is provided as an array, each row of which can be interpreted as the elements of a vector $\mathbf{n}_f$ normal to an internal facet, $f$, which gives rise to a spurious singularity.
The integral is regulated by any vector $\bm{\nu}_\delta$ which obeys $\langle \mathbf{n}_f, \bm{\nu}_\delta \rangle \neq 0 \  \forall f$.

The example \texttt{region\_tools} demonstrates the use of each of the above functions on a 1-loop box integral with an internal mass.

\subsection{New functionalities for coefficients of master integrals}

To evaluate one or several weighted sums of integrals \pysecdec{}
provides the function \py{sum_package()} that takes a list of
integrals $I_i$, and a matrix of coefficients $C_{ki}$ (given
as a list of its rows), so that in the end the weighted sums
$A_k \equiv \sum_i C_{ki} I_i$ are evaluated.
In version~1.5 the coefficients were required to be instances
of the class \py{Coefficient}, and to be specified as a product
of polynomials.

The new version of \pysecdec{} now additionally supports more
flexible ways to specify the coefficient matrix.

\begin{enumerate}
\item
    The coefficients themselves can now be arbitrary arithmetic
    expressions provided as strings. \pysecdec{} now uses
    \textsc{GiNaC}~\cite{Bauer:2000cp} to parse these strings,
    so any syntax recognized by \textsc{GiNaC} is supported.

    The coefficient strings themselves are subsequently used in
    two ways: first during the integral library generation (i.e.
    inside \py{sum_package()}) \pysecdec{} will try to determine
    the leading poles of the coefficients in the regulators, which
    is needed to determine the number of orders the integrals
    will need to be expanded to.
    Second, the strings will be saved to files as they are, and
    loaded back during the evaluation, at which point the symbolic
    variables will be substituted by the values provided by the
    user, and the resulting expressions will be expanded into a
    series in the regulators.
    This evaluation will be performed using arbitrary precision
    rational numbers so that no precision could be lost to numeric
    cancellations.

    This design was chosen to support expressions that are too
    big to be compiled to machine code or to be symbolically
    manipulated in non-trivial ways, such as coefficients arising
    after integration-by-parts reduction.
\item
    Each row of the coefficient matrix can be given either as a
    list of the same size as the number of integrals, or as a
    dictionary from integral indices to coefficients.
    For example, \py{["a","0","b"]} and \py|{0:"a",2:"b"}| are now
    both valid ways to specify the same coefficient matrix row;
    the second way makes it easier to supply sparse matrices
    because zero coefficients can be omitted.
\item
    Each weighted sum can now be given a name.
    To this end, the coefficient matrix can be specified not as
    a list of rows, but rather as a dictionary from sum names
    (i.e. strings) to coefficient matrix rows.
    The supplied names are then used by \textsc{Disteval} in the
    integration log, and in its results, which can optionally be
    structured as dictionaries from the sum name to their values.

    The goal is making it easier to work with multiple sums
    at the same time.
\end{enumerate}

\section{Usage examples and comparison to the previous version}
\label{sec:examples}

The examples described below can be found in the folder {\tt examples/} of the \pysecdec{} distribution.
Unless stated otherwise, the default settings are used.

\subsection{New and featured examples}

We begin by describing the new examples introduced for the current release.
These examples are primarily designed to demonstrate some of the new features.
In \ref{sec:oneloop_ee_mumu} we demonstrate the flexible input syntax for amplitudes and in \ref{sec:twoloop_muon_decay} we show how individual coefficients of the smallness parameter can be accessed when using expansion by regions.
The remaining examples demonstrate the performance of the \Disteval{} and \textsc{IntLib} integrators.

\subsubsection{Simple jupyter notebook examples}

The folder \texttt{examples/jupyter/} contains three examples in the
form of a jupyter notebook where the whole workflow is demonstrated.
These examples are
\begin{description}
\item \texttt{easy.ipynb}: an easy function depending on two parameters;
\item \texttt{box.ipynb}: a one-loop box diagram with massive propagators;
\item \texttt{muon\_production.ipynb}: the one-loop amplitude for
  $e^+e^- \to \mu^+\mu^-$ in massless QED.
\end{description}
Two of the examples are also available without jupyter format, in the
folders \texttt{examples/easy/} and \texttt{examples/muon\_production/}, respectively.

\subsubsection{One-loop amplitude for \texorpdfstring{$e^+e^- \to \mu^+\mu^-$}{e+e- to mu+mu-}}
\label{sec:oneloop_ee_mumu}

The example \texttt{muon\_production} calculates the one-loop
amplitude for muon production in electron-positron annihilation,
$e^+e^- \to \mu^+\mu^-$, with massless leptons in QED. It evaluates a set of scalar master integrals and combines the results with the corresponding integral coefficients. The generation of the amplitude and the Passarino-Veltman reduction of the contributing integrals was done with \textsc{FeynCalc}~\cite{Shtabovenko:2020gxv}. This example is meant to highlight the improved handling of integral coefficients that increases the practicality of using \pysecdec{} for full amplitude calculations.

The \pysecdec{} result for the Born-virtual interference, proportional to $\alpha^3$, where $\alpha$
is the QED fine structure constant, at $s = 3.0$, $t = -1.0$,
$u = -2.0$ (subject to the physical constraint $s+t+u = 0$)
reads\footnote{Here and throughout the paper the numbers in the
parentheses indicate the uncertainty of the final digits. For example,
$1.2345(67)$ means $1.2345\pm0.0067$.}
\begin{equation}
\begin{split}
\mathcal{A}^{(1)}{\mathcal{A}^{(0)}}^*=
& + (-8.704559922781777(7)\cdot 10^{4} + 7(5) \cdot 10^{-11} \, i)\cdot\eps^{-2} \\
& + (+6.1407633077(4) \cdot 10^{4} - 2.73461815073(4) \cdot 10^{5} \, i)\cdot\eps^{-1} \\
& + (+3.45368951804(8) \cdot 10^{5} + 3.98348633939(8) \cdot 10^{5} \, i)\\
& + N_f \big[-2.9015199742604458(3)\cdot 10^{4}\cdot\eps^{-1} \\
& \phantom{{} + N_f \big[} + 3.574514829439898(2) \cdot 10^{4} \\
& \phantom{{} + N_f \big[} - 9.1153938353806605(8) \cdot 10^{4} \, i \big] +\mathcal{O}(\eps)\;,
\end{split}
\end{equation}
where $N_f$ is the number of lepton flavours. The result for the full amplitude has been validated with \textsc{FeynCalc}~\cite{Shtabovenko:2020gxv}. Since the building blocks of this reduced amplitude are only massless integrals, the integration time for one phase space point at the accuracy seen above is in the order of seconds.

\subsubsection{Example from 2-loop muon decay with asymptotic expansion}
\label{sec:twoloop_muon_decay}

\begin{figure}[h]
  \centering
  {\begin{tikzpicture}
	\begin{pgfonlayer}{nodelayer}
		\node [style=none] (0) at (-1, 0) {};
		\node [style=dot] (1) at (0, 0) {};
		\node [style=dot] (2) at (1, 0.75) {};
		\node [style=dot] (3) at (1, -0.75) {};
		\node [style=dot] (4) at (2.5, 0.75) {};
		\node [style=dot] (5) at (2.5, -0.75) {};
		\node [style=none] (6) at (3.5, 0.75) {};
		\node [style=none] (7) at (3.5, -0.75) {};
		\node [style=none] (9) at (0.25, 0.525) {$m_{\tau}$};
		\node [style=none] (10) at (1.725, 1.05) {$M_W$};
		\node [style=none] (11) at (1.75, -0.45) {$M_Z$};
		\node [style=none] (12) at (-0.5, 0.25) {$s$};
	\end{pgfonlayer}
	\begin{pgfonlayer}{edgelayer}
		\draw [style=vector] (0.center) to (1);
		\draw [style=massive edge] (1) to (2);
		\draw [style=vector] (2) to (4);
		\draw [style=vector] (3) to (5);
		\draw [style=outgoing edge] (4) to (6.center);
		\draw [style=outgoing edge] (5) to (7.center);
		\draw [style=edge] (1) to (3);
		\draw [style=edge] (2) to (3);
		\draw [style=edge] (4) to (5);
	\end{pgfonlayer}
\end{tikzpicture}}
  \caption{A 2-loop three point integral with three mass scales.}
  \label{fig:muon_decay2L}
\end{figure}
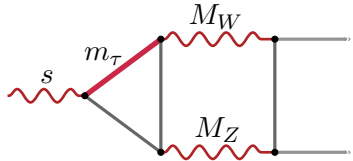

The example \texttt{muon\_decay2L} demonstrates the possibility to produce Python output for each coefficient of an expansion in the smallness parameter within expansion by regions. The diagram in \ref{fig:muon_decay2L} is expanded in the limit of small $\tau$-mass up to order~1, which generates terms with four different powers of $m_{\tau}^2$: $0$, $1$, $1-\eps$ and $1-2\eps$. The result for this diagram reads
\begin{equation}
\begin{split}
  + (m_{\tau}^2)^0 \big[& +(-3.5410(2) + 3.0610(3) \, i) \big] \\
  + (m_{\tau}^2)^1 \big[&+ (-4.93694(1) \cdot 10^{-2} + 2.237604(1) \cdot 10^{-1} \, i)\cdot\eps^{-2} \\
& + (-5.0283(3) \cdot 10^{-1} - 8.7873(3) \cdot 10^{-1} \, i)\cdot\eps^{-1} \\
& + (+2.6476(2) - 1.2090(2) \, i) \big] \\
  + (m_{\tau}^2)^{1-\eps}\big[&+(+9.873890(5)\cdot 10^{-2} - 4.4752040(5) \cdot 10^{-1} \, i)\cdot\eps^{-2} \\
& + (+2.14024(8) \cdot 10^{-1} + 1.97848(7) \cdot 10^{-1} \, i)\cdot\eps^{-1} \\
& + (-7.9370(4) \cdot 10^{-1} + 4.6869(5) \cdot 10^{-1} \, i) \big] \\
  + (m_{\tau}^2)^{1-2\eps}\big[&+ (-4.93694(1)\cdot 10^{-2} + 2.237604(1) \cdot 10^{-1} \, i) \cdot\eps^{-2} \\
& + (+2.8875(3) \cdot 10^{-1} + 6.8082(4) \cdot 10^{-1} \, i)\cdot\eps^{-1} \\
& + (+9.855(1) \cdot 10^{-1} + 2.5875(2) \, i) \big] .
\end{split}
\label{eq:muon_decay_2L_result}
\end{equation}

To obtain the result in this form --- mixing the symbolic prefactors of the form~$(m_{\tau}^2)^k$ with numeric coefficients --- one can generate the integration libraries as in \ref{fig:generate_muon_decay2L_code} and use them for integration as in \ref{fig:integrate_muon_decay2L_code}.
The generation script here is similar to code example~2 in \cite{Heinrich:2021dbf}.
Note that the individual regions in \ref{eq:muon_decay_2L_result} are divergent, however the sum is finite.

On line~4 of \ref{fig:generate_muon_decay2L_code}, \py{LoopIntegralFromGraph()} is used to define a loop integral.
On line~20 this integral is asymptotically expanded in the smallness parameter \texttt{mtsq}${}\equiv m_{\tau}^2$ via the \py{loop_regions()} function up to order~1.
Then, on line~28 the powers of $m_{\tau}^2$ are extracted from the prefactors of the terms of the expansion, and each term has its prefactor modified to no longer include $m_{\tau}$.
On line~31 a mapping between each unique power of the smallness parameter and the corresponding modified terms is added to a dictionary.
Note that several terms may be attributed to the same smallness parameter power.
The final part of the generation script creates the integral libraries corresponding to each unique power of the smallness parameter via the \py{sum_package()} call on line~36.
On line~42 the dictionary mapping powers of the smallness parameter to names of the corresponding integration libraries is saved in a JSON file; this file will later be used by the integration script.

\begin{figure}
\begin{lstlisting}[language=python, numbers=left, xleftmargin=7ex]
import pySecDec as psd, sympy as sp, json

if __name__ == "__main__":
    li = psd.LoopIntegralFromGraph(
        internal_lines = [['mt',[1,4]], ['mw',[4,2]], ['0',[2,3]],
                          ['0',[4,5]], ['0',[1,5]], ['mz',[5,3]]],
        external_lines = [['p1',1], ['p2',2], ['p3',3]],
        regulators = ['eps'],
        replacement_rules = [
            ('p1', '-p2-p3'),
            ('p2*p2', '0'),
            ('p3*p3', '0'),
            ('p2*p3', 's/2'),
            ('mw**2', 'mwsq'),
            ('mz**2', 'mzsq'),
            ('mt**2', 'mtsq')])

    terms = psd.loop_regions(name = 'muon_decay2L',
                             loop_integral = li,
                             smallness_parameter = 'mtsq',
                             decomposition_method = 'geometric',
                             expansion_by_regions_order = 1)

    term_by_prefactor_exponent = {}
    for term in terms:
        coefficient, exponent = sp.sympify(str(term.prefactor)).as_coeff_exponent(sp.sympify('mtsq'))
        term = term._replace(prefactor = coefficient)
        term_by_prefactor_exponent.setdefault(str(exponent), [])
        term_by_prefactor_exponent[str(exponent)].append(term)

    prefactor_exponent_by_name = {}
    for i, (exponent, term) in enumerate(sorted(term_by_prefactor_exponent.items())):
        prefactor_exponent_by_name[f'prefactor_{i+1}'] = exponent
        psd.sum_package(f'prefactor_{i+1}',
                        term,
                        regulators = ['eps'],
                        requested_orders = [0],
                        real_parameters = ['s', 'mwsq', 'mzsq'])

    with open('prefactor_exponent_by_name.json', 'w') as f:
        json.dump(prefactor_exponent_by_name, f)
\end{lstlisting}
    \caption{Generation script for the two-loop muon decay example.}
    \label{fig:generate_muon_decay2L_code}
\end{figure}

\begin{figure}
\begin{lstlisting}[language=python, numbers=left, xleftmargin=7ex]
from pySecDec.integral_interface import DistevalLibrary
import json
import sympy as sp

with open('prefactor_exponent_by_name.json') as f:
    prefactor_exponent_by_name = json.load(f)

result_by_prefactor_exponent = {}
for name, exponent in prefactor_exponent_by_name.items():
    loop_integral = DistevalLibrary(f'{name}/disteval/{name}.json')
    result_by_prefactor_exponent[exponent] = loop_integral(
            parameters = {'s': 3, 'mwsq': 0.78, 'mzsq': 1.0},
            epsrel = 1e-4, points = 1e4, format = 'sympy',
            number_of_presamples = 1e4, timeout = None)

print('Result:')
for exponent, str_result in result_by_prefactor_exponent.items():
    result = sp.sympify(str_result)
    val = result[0].subs({"plusminus": 0})
    err = result[0].coeff("plusminus")
    print(f'''\
  +mtsq^({exponent})*(
    +1/eps^2*(({val.coeff("eps",-2)}) +/- ({err.coeff("eps",-2)}))
    +1/eps^1*(({val.coeff("eps",-1)}) +/- ({err.coeff("eps",-1)}))
    +  eps^0*(({val.coeff("eps",0)}) +/- ({err.coeff("eps",0)}))
  )''')
\end{lstlisting}
    \caption{Integration script for the two-loop muon decay example.}
    \label{fig:integrate_muon_decay2L_code}
\end{figure}

The integration script of \ref{fig:generate_muon_decay2L_code} demonstrates how the \Disteval{} integrator can be called to produce a result of the form given in \ref{eq:muon_decay_2L_result}.
On lines~10 and~11 respectively, each integration library is loaded and called with the kinematic variables $s = 3.0$, $M_W^2 = 0.78$, $M_Z^2 = 1.0$.
Some commonly configured parameters are set explicitly in the library call: \texttt{epsrel} is the relative accuracy, \texttt{points} is the initial QMC lattice size, \texttt{format} is the output format of the result (\py{"sympy"}, \py{"mathematica"}, or \py{"json"}), \texttt{number\_of\_presamples} is the number of samples used for the initial contour deformation parameter selection, and \texttt{timeout} is the maximal allowed integration time in seconds.
The full list of parameters is available in the \pysecdec{} documentation on the \py{DistevalLibrary} class.
The integration script keeps track of which integration library corresponds to which smallness parameter power via the dictionary previously created by the generation script.

\subsubsection{2-loop 5-point hexatriangle example with several mass scales}
\label{sec:hexatriangle}

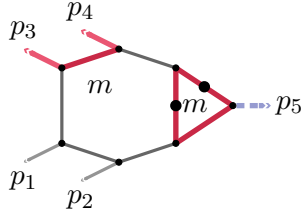
\begin{figure}[h!]
  \centering
  {\begin{tikzpicture}
	\begin{pgfonlayer}{nodelayer}
		\node [style=none] (0) at (-1.25, -0.75) {};
		\node [style=none] (1) at (-0.5, -1) {};
		\node [style=none] (2) at (-1.25, 0.75) {};
		\node [style=none] (3) at (-0.5, 1) {};
		\node [style=none] (4) at (2, 0) {};
		\node [style=dot] (5) at (-0.75, -0.5) {};
		\node [style=dot] (7) at (0, 0.75) {};
		\node [style=dot] (8) at (0.75, 0.5) {};
		\node [style=dot] (9) at (1.5, 0) {};
		\node [style=dot] (10) at (0.75, -0.5) {};
		\node [style=dot] (11) at (0, -0.75) {};
		\node [style=dot] (12) at (-0.75, 0.5) {};
		\node [style=none] (13) at (-1.25, 1) {$p_3$};
		\node [style=none] (14) at (-0.5, 1.25) {$p_4$};
		\node [style=none] (15) at (2.25, 0) {$p_5$};
		\node [style=none] (16) at (-0.25, 0.25) {$m$};
		\node [style=none] (17) at (1, 0) {$m$};
		\node [style=none] (18) at (-1.25, -1) {$p_1$};
		\node [style=none] (19) at (-0.5, -1.25) {$p_2$};
	\end{pgfonlayer}
	\begin{pgfonlayer}{edgelayer}
		\draw [style=outgoing edge] (11) to (1.center);
		\draw [style=outgoing edge] (5) to (0.center);
		\draw [style=outgoing massive edge] (7) to (3.center);
		\draw [style=outgoing massive scalar] (9) to (4.center);
		\draw [style=edge] (5) to (12);
		\draw [style=massive edge] (12) to (7);
		\draw [style=edge] (7) to (8);
		\draw [style=massive edge, style=dot1] (8) to (9);
		\draw [style=massive edge] (9) to (10);
		\draw [style=massive edge, style=dot1] (10) to (8);
		\draw [style=edge] (10) to (11);
		\draw [style=edge] (11) to (5);
		\draw [style=outgoing massive edge] (12) to (2.center);
	\end{pgfonlayer}
\end{tikzpicture}}
  \caption{A 2-loop 5-point integral with massive propagators and
    massive legs. The integral is evaluated in $6-2\eps$ space-time
    dimensions. The configuration being tested is
    $p_1^2=p_2^2=0$,
    $p_3^2=p_4^2=m^2=1$,
    $p_5^2 = 12/23$,
    $(p_1+p_2)^2 = 262/35$,
    $(p_2+p_3)^2 = -92/53$,
    $(p_3+p_5)^2 = 491/164$,
    $(p_5+p_4)^2 = 373/124$,
    $(p_4+p_1)^2 = -65/36$.}
  \label{fig:hexatriangle}
\end{figure}

The example \texttt{hexatriangle} is a 2-loop 5-point integral
depicted in \ref{fig:hexatriangle}. This is a master integral
for the amplitude of $q\bar{q} \to t\bar{t}H$ production at
two loops. The integral is dimensionally shifted to $6-2\eps$
space-time dimensions; the dimensional shift and additional dots
were chosen to make it finite in~$\eps$ and fast to evaluate.

\begin{figure}
  \centering
  \includegraphics[width=\textwidth]{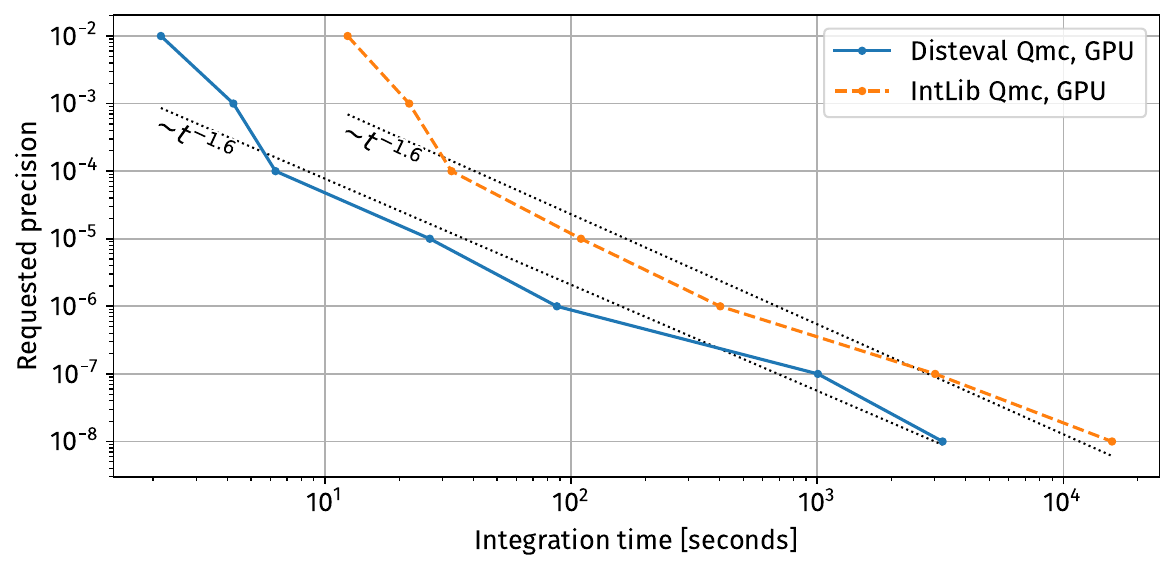}
  \caption{
      The obtained precision by integration time for the \texttt{hexatriangle} example.
      This plot is based on the data from \ref{tab:hexatriangle-timings}.
      \label{fig:hexatriangle-timing-plot}}
\end{figure}

The value of the integral at the point specified in \ref{fig:hexatriangle} is
\begin{equation}
    1.454919812(7) \cdot 10^{-7} - 1.069797219(8) \cdot 10^{-7}\,i + \mathcal{O}(\eps).
\end{equation}
The convergence rate of the integral is depicted in \ref{fig:hexatriangle-timing-plot}.
Overall the obtained precision scales with the integration time~$t$ approximately as~$1/t^{1.6}$.
We want to emphasise that such scaling is made possible by the
use of the QMC integration methods; traditional
Monte Carlo methods only scale as fast as~$1/t^{0.5}$.

A more detailed list of integration timings is given in \ref{tab:hexatriangle-timings}.

\begin{table}
    \centering
    \scalebox{0.95}{%
        \begin{tabular}{rlrrrrrrr}
        \toprule
        \multicolumn{2}{r}{\textsubscript{Integrator}\textbackslash\textsuperscript{Accuracy}} & $10^{-2}$ & $10^{-3}$ & $10^{-4}$ & $10^{-5}$ & $10^{-6}$ & $10^{-7}$ & $10^{-8}$ \\
        \midrule
        \midrule
        GPU & \textsc{Disteval} &  2.2\,s &  4.2\,s &  6.3\,s &  27\,s & 1.5\,m & 17\,m &  54\,m \\
            & \textsc{IntLib}   & 12.3\,s & 22.0\,s & 32.6\,s & 110\,s & 6.7\,m & 50\,m & 263\,m \\
            & Ratio             &     5.6 &     5.2 &     5.2 &    4.1 &    5.6 &   3.0 &    4.9 \\
        \midrule
        CPU & \textsc{Disteval} & 3.5\,s &  5.1\,s & 14\,s &  1.6\,m &  8.3\,m &  57\,m &  4.7\,h \\
            & \textsc{IntLib}   & 8.5\,s & 20.8\,s & 86\,s & 14.2\,m & 62.2\,m & 480\,m & 43.1\,h \\
            & Ratio             &    2.4 &     4.1 &   6.1 &     8.7 &    7.5 &     8.4 &     9.2 \\
        \bottomrule
        \end{tabular}%
    }
    \caption{
        Integration timings for the \texttt{hexatriangle} example
        (\ref{fig:hexatriangle}) depending on the requested
        accuracy using two integrators: \textsc{Disteval} and
        \textsc{IntLib} \texttt{Qmc}.
        The GPU timings were taken using an NVidia~A100~80G,
        with the integrands compiled using \textsc{Cuda}~11.8.89.
        The CPU timings are for an AMD~\textsc{Epyc}~7F32
        processor with 16~cores (32~threads), with the integrands
        compiled using GCC~12.2.1 with \texttt{CXXFLAGS} set to
        \texttt{-O3~-mavx2~-mfma}.
        \label{tab:hexatriangle-timings}
    }
\end{table}

\subsubsection{2-loop 5-point offshell pentabox example}
\label{sec:pentabox-offshell}

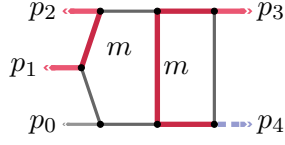
\begin{figure}[h]
  \centering
  {\begin{tikzpicture}
	\begin{pgfonlayer}{nodelayer}
		\node [style=none] (0) at (-1.25, -0.75) {};
		\node [style=none] (1) at (-1.5, 0) {};
		\node [style=none] (2) at (-1.25, 0.75) {};
		\node [style=none] (3) at (1.25, 0.75) {};
		\node [style=none] (4) at (1.25, -0.75) {};
		\node [style=dot] (5) at (-0.75, -0.75) {};
		\node [style=dot] (6) at (-1, 0) {};
		\node [style=dot] (7) at (-0.75, 0.75) {};
		\node [style=dot] (8) at (0.75, 0.75) {};
		\node [style=dot] (9) at (0.75, -0.75) {};
		\node [style=dot] (10) at (0, -0.75) {};
		\node [style=dot] (11) at (0, 0.75) {};
		\node [style=none] (13) at (0.25, 0) {$m$};
		\node [style=none] (15) at (-1.5, -0.75) {$p_0$};
		\node [style=none] (16) at (-1.75, 0) {$p_1$};
		\node [style=none] (17) at (-1.5, 0.75) {$p_2$};
		\node [style=none] (18) at (1.5, 0.75) {$p_3$};
		\node [style=none] (19) at (1.5, -0.75) {$p_4$};
		\node [style=none] (20) at (-0.5, 0.25) {$m$};
	\end{pgfonlayer}
	\begin{pgfonlayer}{edgelayer}
		\draw [style=outgoing edge] (5) to (0.center);
		\draw [style=outgoing massive edge] (6) to (1.center);
		\draw [style=outgoing massive edge] (7) to (2.center);
		\draw [style=outgoing massive edge] (8) to (3.center);
		\draw [style=outgoing massive scalar] (9) to (4.center);
		\draw [style=edge] (10) to (5);
		\draw [style=edge] (5) to (6);
		\draw [style=edge] (7) to (11);
		\draw [style=edge] (8) to (9);
		\draw [style=massive edge] (9) to (10);
		\draw [style=massive edge] (10) to (11);
		\draw [style=massive edge] (11) to (8);
		\draw [style=massive edge] (6) to (7);
	\end{pgfonlayer}
\end{tikzpicture}}
  \caption{A 2-loop 5-point pentabox integral with massive propagators and
    massive legs. The configuration being tested is
    $p_0^2=0$,
    $p_1^2=p_2^2=p_3^2=m^2=1/2$,
    $(p_0+p_1)^2 = 2.2$,
    $(p_0+p_2)^2 = 2.3$,
    $(p_0+p_3)^2 = 2.4$,
    $(p_1+p_2)^2 = 2.5$,
    $(p_1+p_3)^2 = 2.6$,
    $(p_2+p_3)^2 = 2.7$.}
  \label{fig:pentabox-offshell}
\end{figure}

The example \texttt{pentabox\_offshell} is an integral depicted
in \ref{fig:pentabox-offshell}. It is a 2-loop pentabox with
an internal mass, massive legs, and the total of 7 scales. The
integral is evaluated in $6-2\eps$ space-time dimensions (where
it is finite in $\eps$) up to $\mathcal{O}(\eps^4)$; a prefactor
of $\Gamma(2+2\eps)$ is divided out to match the configuration of
Section~6.4 of~\cite{Borinsky:2023jdv}, where the same integral
is calculated numerically via tropical integration.

The value of the integral at the point specified in \ref{fig:pentabox-offshell} is
\begin{equation}
\begin{split}
    &+(+6.443869(7) \cdot 10^{-2}-8.267759(7) \cdot 10^{-2} \, i) \, \eps^0 \\
    &+(+4.043397(2) \cdot 10^{-1}+3.189607(2) \cdot 10^{-1} \, i) \, \eps^1 \\
    &+(-7.771389(2) \cdot 10^{-1}+9.370171(2) \cdot 10^{-1} \, i) \, \eps^2 \\
    &+(-1.3220709(6) \cdot 10^{0}-1.2139678(6) \cdot 10^{0} \, i) \, \eps^3 \\
    &+(+1.3789155(10) \cdot 10^{0}-1.2118956(10) \cdot 10^{0} \, i) \, \eps^4 \\
    &+\mathcal{O}(\eps^5)
\end{split}
\end{equation}

These values match the ones given in~\cite{Borinsky:2023jdv} within the uncertainty limits.

\begin{figure}
  \centering
  \includegraphics[width=\textwidth]{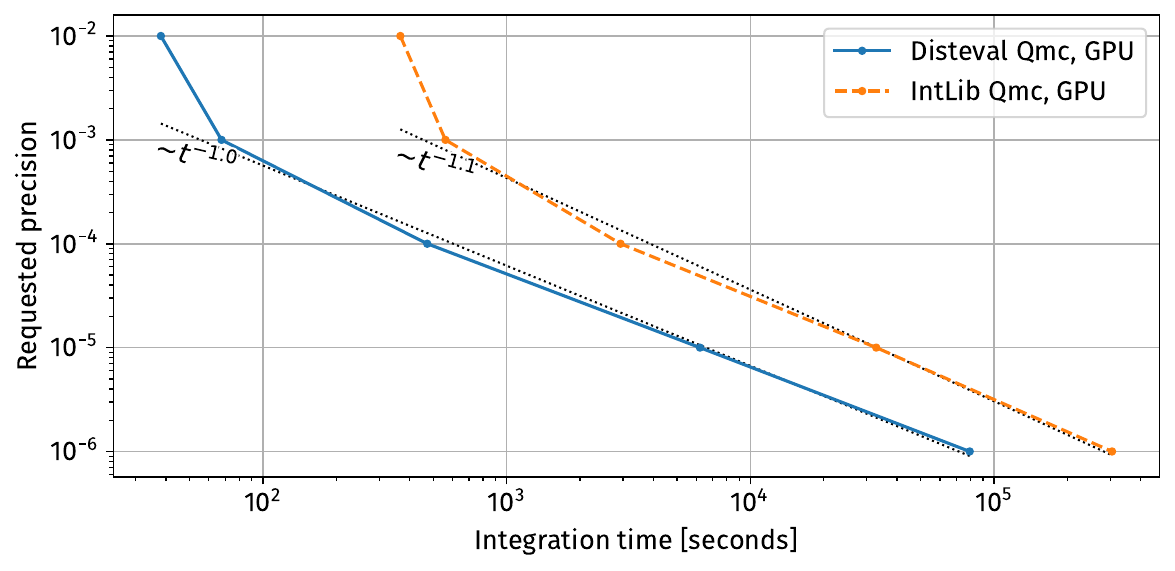}
  \caption{
      The obtained precision by integration time for the \texttt{pentabox\_offshell} example.
      This plot is based on the data from \ref{tab:pentabox-offshell-timings}.
      \label{fig:pentabox-offshell-timing-plot}}
\end{figure}

The convergence rate of the integral is depicted in \ref{fig:pentabox-offshell-timing-plot}.
Overall the obtained precision scales with the integration time~$t$ approximately as~$1/t$.

A more detailed list of integration timings is given in \ref{tab:pentabox-offshell-timings}.

\begin{table}
    \centering
    \scalebox{0.95}{%
        \begin{tabular}{rlrrrrrrr}
        \toprule
        \multicolumn{2}{r}{\textsubscript{Integrator}\textbackslash\textsuperscript{Accuracy}} & $10^{-2}$ & $10^{-3}$ & $10^{-4}$ & $10^{-5}$ & $10^{-6}$ \\
        \midrule
        \midrule
        GPU & \textsc{Disteval} &  38\,s & 1.1\,m &  7.9\,m & 1.7\,h & 22\,h \\
            & \textsc{Intlib}   & 366\,s & 9.3\,m & 48.9\,m & 9.1\,h & 85\,h \\
            & Ratio             &    9.6 &    8.3 &     6.2 &    5.3 &   3.8 \\
        \midrule
        CPU & \textsc{Disteval} & 13\,s &  2.4\,m &  43\,m &  7.9\,h & --- \\
            & \textsc{Intlib}   & 67\,s & 18.9\,m & 299\,m & 65.0\,h & --- \\
            & Ratio             &   5.0 &     7.8 &    7.0 &     8.2 & --- \\
        \bottomrule
        \end{tabular}%
    }
    \caption{
        Integration timings for the \texttt{pentabox\_offshell}
        example (\ref{fig:pentabox-offshell}) depending on the
        requested accuracy using two integrators: \textsc{Disteval}
        and \textsc{IntLib} \texttt{Qmc}.
        Same benchmarking conditions as in \ref{tab:hexatriangle-timings}.
        \label{tab:pentabox-offshell-timings}
    }
\end{table}

\subsubsection{4-loop triangle diagram}
\label{sec:fourloop_triangle}

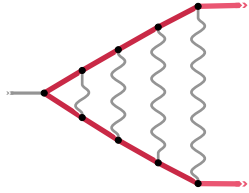
\begin{figure}[h]
  \centering
  {\begin{tikzpicture}
	\begin{pgfonlayer}{nodelayer}
		\node [style=none] (0) at (-1.25, 0) {};
		\node [style=dot] (1) at (-0.75, 0) {};
		\node [style=dot] (2) at (-0.25, 0.3) {};
		\node [style=dot] (3) at (-0.25, -0.325) {};
		\node [style=dot] (4) at (0.225, 0.575) {};
		\node [style=dot] (5) at (0.225, -0.625) {};
		\node [style=dot] (6) at (0.75, 0.875) {};
		\node [style=dot] (7) at (0.75, -0.925) {};
		\node [style=dot] (8) at (1.275, 1.15) {};
		\node [style=dot] (9) at (1.275, -1.2) {};
		\node [style=none] (10) at (1.925, 1.165) {};
		\node [style=none] (11) at (1.925, -1.205) {};
	\end{pgfonlayer}
	\begin{pgfonlayer}{edgelayer}
		\draw [style=incoming edge] (0.center) to (1);
		\draw [style=massive edge] (1) to (2);
		\draw [style=massive edge] (2) to (4);
		\draw [style=massive edge] (4) to (6);
		\draw [style=massive edge] (6) to (8);
		\draw [style=massive edge] (1) to (3);
		\draw [style=massive edge] (3) to (5);
		\draw [style=massive edge] (5) to (7);
		\draw [style=massive edge] (7) to (9);
		\draw [style=outgoing massive edge] (9) to (11.center);
		\draw [style=outgoing massive edge] (8) to (10.center);
		\draw [style=incoming vector] (3) to (2);
		\draw [style=incoming vector] (4) to (5);
		\draw [style=incoming vector] (6) to (7);
		\draw [style=incoming vector] (9) to (8);
	\end{pgfonlayer}
\end{tikzpicture}}
  \caption{A 4-loop diagram with kinematics inspired by  contributions
    to the electron or muon anomalous magnetic moment.}
  \label{fig:gminus2_4L}
\end{figure}

The example \texttt{gminus2\_4L} is a four-loop diagram
contributing to the electron or muon anomalous magnetic moment.
The diagram is depicted in \ref{fig:gminus2_4L}.
The massive lines (coloured in red) denote on-shell massive fermion
lines, $p^2=m^2$. For the grey external line with momentum $q$, the
limit $q\to 0$ needs to be taken, such that the diagram is
characterised by $q^2=0, q\cdot p=0, p^2=m^2$. Therefore the
corresponding integral becomes a single scale integral, depending only
on $m^2$.

The \pysecdec{} result for \texttt{gminus2\_4L} reads
\begin{equation}
\begin{split}
&+ 2.60420(2) \cdot 10^{-3} \cdot \eps^{-4} \\
& + 2.5237(2) \cdot 10^{-2} \cdot \eps^{-3} \\
& + 3.8721(4) \cdot 10^{-1} \cdot \eps^{-2} \\
& + 3.9116(4) \cdot \eps^{-1} \\
& + 39.256(4) + \mathcal{O}(\eps).
\end{split}
\end{equation}

\subsubsection{6-loop two-point function}
\label{sec:6-loop two-point function}

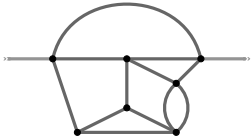
\begin{figure}[h]
  \centering
  {\begin{tikzpicture}[scale=1.3]
	\begin{pgfonlayer}{nodelayer}
		\node [style=none] (0) at (-0.5, 0) {};
		\node [style=none] (1) at (2, 0) {};
		\node [style=dot] (3) at (0, 0) {};
		\node [style=dot] (4) at (1.5, 0) {};
		\node [style=dot] (5) at (1.25, -0.25) {};
		\node [style=dot] (6) at (0.75, 0) {};
		\node [style=dot] (7) at (0.75, -0.5) {};
		\node [style=dot] (8) at (1.25, -0.75) {};
		\node [style=dot] (10) at (0.25, -0.75) {};
	\end{pgfonlayer}
	\begin{pgfonlayer}{edgelayer}
		\draw [style=edge] (3) to (10);
		\draw [style=edge] (10) to (7);
		\draw [style=edge] (7) to (8);
		\draw [style=edge, bend left=45] (8) to (5);
		\draw [style=edge] (5) to (6);
		\draw [style=edge] (6) to (3);
		\draw [style=edge] (6) to (4);
		\draw [style=edge] (6) to (7);
		\draw [style=edge] (8) to (10);
		\draw [style=edge, bend left=75, looseness=1.25] (3) to (4);
		\draw [style=edge] (4) to (5);
		\draw [style=outgoing edge] (4) to (1.center);
		\draw [style=incoming edge] (0.center) to (3);
		\draw [style=edge, bend right=45] (8) to (5);
	\end{pgfonlayer}
\end{tikzpicture}}
  \caption{A 6-loop two-point integral.}
  \label{fig:6Lbubble}
\end{figure}

The \texttt{bubble6L} example consists of the 6-loop 2-point integral shown in \ref{fig:6Lbubble}. The pole coefficients are given
analytically in Eq.~(A3) of Ref.~\cite{Kompaniets:2017yct} (at $p^2=-p_E^2=-1$, where $p_E$ is the external momentum in Euclidean space).
%A prefactor of $\left(\frac{\Gamma(1-\eps)^2\Gamma(1+\eps)}{\Gamma(2-2\eps)}\right)$ per loop is extracted in Ref.~\cite{Kompaniets:2017yct}.
%Note that a factor of $\Gamma(6\eps)$ comes from the Feynman parametrisation and is included in the numerical result.
%The \pysecdec{} symmetry finder reduces the number of sectors from more than 14000 to 8774.
%
The decomposition method {\tt `geometric'} is the default and recommended decomposition method in version 1.6.
In this example it is mandatory to use a geometric decomposition because  {\tt `iterative'}  leads to an infinite recursion.
Usually the geometric decomposition method produces the fewest sectors.
However, for graphs with very high symmetry, the iterative method can occasionally produce fewer sectors than the geometric method as it does not destroy symmetries when one of the Feynman parameters is eliminated using the $\delta$-constraint.
More information about the various decomposition methods can be found in Refs.~\cite{Borowka:2015mxa,Borowka:2017idc,Schlenk:2016cwf} and in the code documentation.
%(section 2.2.1 of the manual).

The analytic result is given by
\begin{align}
B_{6L}^{\text{analyt.}}&= \frac{1}{\eps^2}\,\frac{147}{16}\,\zeta_7
- \frac{1}{\eps}\,\left(\frac{147}{16}\,\zeta_7 +\frac{27}{2}\,\zeta_3\zeta_5+\frac{27}{10}\zeta_{3,5}-\frac{2063}{504000}\,\pi^8\right) \;+\;\mathcal{O}(\eps^0)\nn\\
&= \frac{9.264208985946416}{\eps^2} + \frac{91.73175282208716}{\eps}  \;+\;\mathcal{O}(\eps^0) \;.
\end{align}

The \pysecdec{} result at $p^2=-1$ obtained with the \Disteval{} integrator reads
\begin{equation}
\begin{split}
B_{6L}^{\text{num.}}=
& + 9.26420902(3) \cdot\eps^{-2} \\
& + 9.17317528(8)\cdot 10^{1} \cdot\eps^{-1} \\
&+ 1.11860698(1)\cdot10^{3}+\mathcal{O}(\eps)\;.
\end{split}
\end{equation}

\subsection{Previously existing examples}
\label{sec:timings}

\begin{figure}[]
\centering
\begin{subfigure}[b]{0.3\textwidth}
         \centering
         \sfig{figures/banana3mass.tikz}
         \caption{\texttt{banana\_3mass}}
         \label{fig:banana3mass}
     \end{subfigure}
\begin{subfigure}[b]{0.3\textwidth}
         \centering
         \sfig{figures/pentabox.tikz}
         \caption{\texttt{pentabox\_fin}}
         \label{fig:pentabox}
     \end{subfigure}
\begin{subfigure}[b]{0.3\textwidth}
         \centering
         \sfig{figures/formfactor4L.tikz}
         \caption{\texttt{formfactor4L}}
         \label{fig:formfactor4L}
     \end{subfigure}
\begin{subfigure}[b]{0.3\textwidth}
         \centering
         \sfig{figures/hz2L_nonplanar.tikz}
         \caption{\texttt{hz2L\_nonplanar}}
         \label{fig:hz2L_nonplanar}
     \end{subfigure}
\begin{subfigure}[b]{0.3\textwidth}
         \centering
         \sfig{figures/elliptic2L.tikz}
         \caption{\texttt{elliptic2L\_physical}}
         \label{fig:elliptic2L}
     \end{subfigure}
\begin{subfigure}[b]{0.3\textwidth}
         \centering
         \sfig{figures/Nbox2L_split_b.tikz}
         \caption{\texttt{Nbox2L\_split\_b}}
         \label{fig:Nbox}
     \end{subfigure}
\caption{All diagrams of \ref{tab:timings_pysd} except for \texttt{bubble6L}, which is described in detail in \ref{sec:6-loop two-point function}.}
    \label{fig:timings_diagrams}
\end{figure}

% \begin{table}[htb]
% \begin{center}
% \begin{tabular}{|l|c|c|c|c|}
% \hline
% & &v1.5.6 (\qmc{})&  v1.6 (disteval \qmc{}) \\
% & rel. acc. & time\,[s] & time\,[s] \\
% \hline
% \scriptsize \texttt{banana 3mass 3L}      & $10^{-9}$ & 29.5 &  9.8  \\
% \scriptsize \texttt{HZ nonplanar 2L}      & $10^{-4}$ & 140.5 & 21.4\\
% \scriptsize \texttt{pentabox fin 2L}      & $10^{-4}$ & 229.5 & 32.5\\
% \scriptsize \texttt{elliptic 2L}          & $10^{-4}$ & 15.6 &  8.4\\
% \scriptsize \texttt{formfactor 4L}        & $10^{-4}$ & 9379 & 505.1\\
% \scriptsize \texttt{Nbox split b 2L}      & $10^{-4}$ & 538.7& 89.4\\
% \scriptsize \texttt{bubble 6L}            & $10^{-4}$ & 4005.9 & 222.7\\
% \hline
% \end{tabular}
% \end{center}
% \caption{Comparison of timings using the \qmc{} integrator of
%   \pysecdec{} version 1.5.6 versus the disteval \qmc{} integrator of this release.
% The obtained relative accuracy refers to the finite real part
% of the integral including all prefactors.\label{tab:timings_pysd}}
% \end{table}

\begin{table}
    \resizebox{\columnwidth}{!}{%
    \begin{tabular}{rlrrrrrrr}
    \toprule
    \multicolumn{2}{r}{\textsubscript{Integrator}\textbackslash\textsuperscript{Accuracy}} & $10^{-2}$ & $10^{-3}$ & $10^{-4}$ & $10^{-5}$ & $10^{-6}$ & $10^{-7}$ & $10^{-8}$ \\
    \midrule
    \midrule
    \texttt{banana\_3mass}
        & \textsc{Disteval} & 2.1\,s & 2.1\,s & 2.4\,s & 2.6\,s & 2.6\,s & 2.9\,s &  3.6\,s \\
        & \textsc{IntLib}   & 5.0\,s & 4.9\,s & 6.4\,s & 7.2\,s & 8.5\,s & 8.5\,s & 13.8\,s \\
        & Ratio             &    2.3 &    2.3 &    2.7 &    2.7 &    3.2 &    3.0 &     3.9 \\
    \midrule
    \texttt{bubble6L}
        % & \textsc{Disteval} &  1.7\,m &  1.8\,m &  1.8\,m &  2.0\,m &  3.5\,m &   9.5\,m &  1.2\,h \\
        & \textsc{Disteval} &  1.8\,m &  1.8\,m &  1.8\,m &  2.1\,m &  3.8\,m &  10.2\,m &  1.2\,h \\
        & \textsc{IntLib}   & 39.5\,m & 38.8\,m & 39.6\,m & 43.8\,m & 85.1\,m  & 170.7\,m & 11.6\,h \\
        & Ratio             &      22 &      22 &      22 &      21 &      22 &       17 &      10 \\
    \midrule
    \texttt{formfactor4L}
        % & \textsc{Disteval} & 3.6\,m &  3.7\,m &  3.7\,m &  3.7\,m &  6.2\,m &  0.21\,h & 0.91\,h \\
        & \textsc{Disteval} &  4.1\,m &  4.1\,m &  4.1\,m &  4.4\,m &  7.7\,m &  14.6\,m &  0.96\,h \\
        & \textsc{IntLib}   & 74\,m & 73\,m & 73\,m & 74\,m & 136\,m & 246\,m & 10.9\,h \\
        & Ratio             &   18  &    18 &     18 &      17 &    18 &     17  &   11 \\
    \midrule
    \texttt{elliptic2L\_physical}
        & \textsc{Disteval} & 1.6\,s & 1.5\,s & 1.7\,s & 1.9\,s &  4.0\,s & 19\,s & 7.6\,m \\
        & \textsc{IntLib}   & 3.1\,s & 4.8\,s & 4.9\,s & 7.3\,s & 13.8\,s & 53\,s & 4.3\,m \\
        & Ratio             &    1.9 &    3.1 &    2.8 &    3.9 &     3.4 &   2.9 &    0.6 \\
      \midrule
  \texttt{hz2L\_nonplanar}
        % & \textsc{Disteval} & 5\,s &  5\,s &  9\,s &  37\,s &  2.3\,m &  5.4\,m & 27.1\,m \\
        & \textsc{Disteval} &  2.1\,s &  2.6\,s &  4.6\,s &  30.4\,s &  2.2\,m &  5.1\,m &  27.1\,m \\
        & \textsc{IntLib}   & 9\,s & 17\,s & 41\,s & 163\,s & 9.6\,m & 16.0\,m & 27.3\,m \\
        & Ratio             &  1.8   &    3.4 &     4.6 &      4.4 &    4.2 &     3.0 &   1.0 \\
    \midrule
    \texttt{Nbox2L\_split\_b}
        % & \textsc{Disteval} & 8\,s &  16\,s &  23\,s &  40\,s &  2.4\,m  &  9.1\,m & 19.9\,m \\
        & \textsc{Disteval} &  2.7\,s &  9.8\,s &  16.8\,s &  0.58\,m &  2.4\,m &  9.1\,m &  20\,m \\
        & \textsc{IntLib}   & 24\,s & 73\,s & 223\,s & 6.6\,m & 26\,m & 43\,m & 93\,m \\
        & Ratio             &   3.0  &    4.6 &     9.7 &      9.9 &    10.5 &     4.8 &   4.7 \\
    \midrule
    \texttt{pentabox\_fin}
        & \textsc{Disteval} &  5\,s &  8\,s & 11\,s  & 0.71\,m &  3.7\,m & 18.5\,m & 1.1\,h \\
        & \textsc{IntLib}   & 45\,s & 65\,s & 88\,s &  3.2\,m & 11.3\,m & 74.8\,m & 4.6\,h \\
        & Ratio             &   8.6 &   7.9 &   7.7 &     4.5 &     3.1 &     4.0 &    4.2 \\
    \bottomrule
    \end{tabular}%
    }
    \caption{Integration timings on a GPU for different examples
    using the \textsc{IntLib} \texttt{Qmc} integrator and the
    new \textsc{Disteval} integrator. All timings are using the CBC
    generating vectors from the previous release, meaning the ratios
    between \textsc{IntLib} and \textsc{Disteval} are purely due to
    the improvements described in \ref{sec:disteval}.  The
    significantly improved timings achieved by using the new median
    generating vectors are shown in~\ref{tab:elliptic2L_physical-median-timings}.} 
    \label{tab:timings_pysd}
\end{table}
% \begin{figure}[]
% \centering
% \begin{subfigure}[b]{\textwidth}
%          \centering
%          \includegraphics[width=\textwidth]{figures/light_benchmarks_convergence.pdf}
%          \caption{Lighter benchmarks}
%          \label{fig:light_convergence}
%      \end{subfigure}
% \begin{subfigure}[b]{\textwidth}
%          \centering
%          \includegraphics[width=\textwidth]{figures/heavy_benchmarks_convergence.pdf}
%          \caption{Heavier benchmarks}
%          \label{fig:heavy_convergence}
%      \end{subfigure}
% \caption{Convergence rates of the examples from \ref{tab:timings_pysd}.}
% \end{figure}
\begin{figure}[]
    \centering
    \includegraphics[width=\textwidth]{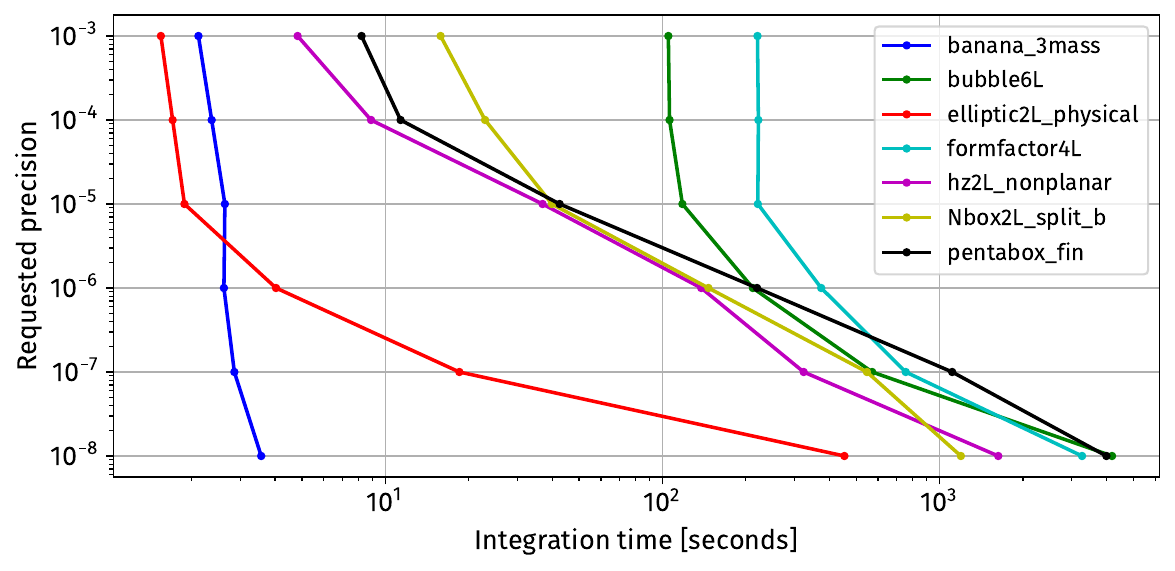}
    \caption{Convergence rates of the \Disteval{} timings from \ref{tab:timings_pysd}}
    \label{fig:disteval-timings}
\end{figure}

Several previously existing \pysecdec{} examples, shown in \ref{fig:timings_diagrams}, have been benchmarked in~\cite{Borowka:2018goh}.
In \ref{tab:timings_pysd} and \ref{fig:disteval-timings} we provide a comparison of the integration time of those examples using the \textsc{Disteval} integrator (new in v1.6) and the \textsc{IntLib} \texttt{Qmc} integrator (the default of v1.5.6), all on an NVidia~A100~80G GPU (using \textsc{Cuda} version 11.8).

The reported integration times correspond to the
wall clock times for running the integration via the Python interface
of \pysecdec{}. In particular, the numerical integration of \emph{all} orders in $\eps$ up to the finite order
is included in the timings. The precision refers to the relative error which in this case is defined as $\epsilon_{\text{rel}}~=~\sqrt{\frac{(\Delta R)^2+(\Delta I)^2}{R^2 + I^2}}$, $R$ and $I$ are the real and imaginary parts of a coefficient in the $\eps$-expansion, and $\Delta R$ and $\Delta I$ are the corresponding uncertainties.
The examples \texttt{formfactor4L}  and \texttt{bubble6L} have been calculated using the \texttt{baker} integral transformation, for the other examples the default transformation \texttt{korobov3} has been used.
% These two examples have also been evaluated for a dynamic number of \texttt{QMC} lattice points, depending on the requested accuracy. For accuracies $10^{-2}$, $10^{-3}$ and $10^{-4}$, the number of lattice points was $100$, $500$ and $1000$ respectively. For the remaining higher accuracies, the number of points was fixed to 10000.

The overall conclusion is that \textsc{Disteval} is 3$\times$-5$\times$ faster
than \textsc{IntLib} \texttt{Qmc} with equivalent settings on a
GPU at higher accuracies, with the exception of the Euclidean
integrals \texttt{bubble6L} and \texttt{formfactor4L}. They
contain a large number of sectors, each very simple, so that
the execution time is mostly dominated by overhead. 
\textsc{Disteval} has up to 20$\times$ less overhead.

Of particular note is the benchmark of the \texttt{elliptic2L\_physical}
example: at the requested precision of $10^{-8}$, the speedup of
\textsc{Disteval} is 0.6, so it is slower than \textsc{IntLib}
\texttt{Qmc}.
The reason for this is exactly the unlucky lattice at
$n=4.3\cdot\!10^9$ depicted in \ref{fig:precision-by-lattice}:
\textsc{Disteval} reaches it first at this requested precision,
while \textsc{IntLib} does not hit this particular lattice
because its algorithm of selecting $n_i$ differs just slightly
enough to land on a nearby lattice instead.
In any case, this problem is circumvented by the use of median QMC
rules, and we have investigated the \texttt{elliptic2L\_physical}
example in detail in \ref{sec:medianlattice}.

%\clearpage

\section{Conclusions}
\label{sec:conclusion}
We have presented version 1.6 of \pysecdec{}, featuring a major
upgrade targeted at the evaluation of loop amplitudes through a novel, highly distributed Quasi-Monte-Carlo (QMC) evaluation method.
Compared to the previous version, the virtues of the new method applied to individual multi-loop integrals are particularly manifest for multi-scale integrals and when high precision is requested. 
Very importantly, the calculation of {\em amplitudes} rather than individual integrals is facilitated.
This is achieved  through several improvements, for instance, new functionalities to treat the coefficients of master integrals, which are typically large expressions after IBP reduction. 
Furthermore,  amplitudes are calculated as weighted sums of integrals with coefficients, with an overall precision goal that can be specified by the user.
A new integrator based on median QMC rules avoids the limitations of the component-by-component construction of generating vectors for lattice rules. 
It also remedies the intermediate loss of QMC-typical scaling that has been observed for some fixed individual lattices.

The release contains improvements to the expansion by regions functionality, including  the  treatment of integrals
with numerators within expansion by regions and the automated detection of whether and where additional regulators are needed, making this information completely transparent to the user.
The coefficients of each order of the expansion in the small parameter are now also easily accessible to the user.

With these new features \pysecdec{} is significantly faster, more flexible, and easier to use than previous versions.
It is better equipped to analyse and tackle a wide range of problems including previously intractable multi-loop amplitudes needed for precision phenomenology, problems requiring multiple dimensional regulators, and integrals/amplitudes where higher numerical precision than previously possible is required.

\section*{Acknowledgements}

We would like to thank Goutam Das, Joshua Davies, Christoph Greub, Andrey Pikelner, Vladyslav Shtabovenko and Yannick Ulrich for discussions and raising issues that helped to improve the program.
This research was supported by  the  Deutsche  Forschungsgemeinschaft (DFG, German Research Foundation) under grant 396021762 - TRR 257.
SJ is supported by a Royal Society University Research Fellowship (Grant URF/R1/201268).

\renewcommand \thesection{\Alph{section}}
\appendix
\setcounter{section}{0}
\setcounter{equation}{0}
%\input{appendix}

%\section*{References}
%\addcontentsline{toc}{chapter}{References}

\bibliographystyle{JHEP}

\renewcommand*{\bibfont}{\justify}
\bibliography{deval}

\end{document}